Surface slopes of asteroid pairs as indicators of mechanical properties and cohesion

David Polishook, Oded Aharonson

**Abstract**

Asteroid pairs had a single progenitor that split due to rotational-fission of a weak, rubble-pile structured body. By constructing shape models of asteroid pairs from multiple-apparition observations and using a lightcurve inversion technique, we mapped the gravitational and rotational accelerations on the surfaces of these asteroids. This allows us to construct a map of local slopes on the asteroids' surfaces. In order to test for frictional failure, we determine the maximum rotation rate at which an area larger than half the surface area of the secondary member (assumed to be the ejected component) has a slope value greater than 40 degrees, the angle of friction of lunar regolith, where loose material will begin sliding. We use this criterion to constrain the failure stress operating on the body, just before disruption at the commonly observed spin barrier of 2.2 h.

Our current sample includes shape models of eleven primary members of asteroid pairs, observed from the Wise Observatory in the last decade. In the studied parameter space we find that the shape models only reach the spin barrier when their bulk density is larger than the $\sim2$ gr cm$^{-3}$ measured for the rubble pile structured 25143 Itokawa, suggesting that km-sized asteroid pairs are dense compared to sub-km bodies. Assuming ejection of secondary components that are larger than those observed (up to the maximal size allowing separation), can also increase the spin barrier of the asteroids, thus supporting the previously suggested scenario of continuous disruption of the secondary. In addition, cohesion levels of hundreds of Pascals are also required to prevent these shape models from disrupting at spin rates slower than the usual spin barrier.

**Introduction**

In recent years it became clear that a significant number of asteroids have a fragile nature. Various measurements provide evidence that most asteroids in the size range of $\sim0.2$ to tens of km are collections of aggregates separated by voids, otherwise known as *rubble piles*: Asteroid densities are lower than the density of their constituent elements (e.g. Fujiwara et al. 2006; Carry 2012); Almost all asteroids larger than 0.2 km rotate slower than $\sim2.2$ h per cycle (the *rubble pile spin barrier*) suggesting their internal strength is too weak to sustain a fast rotation (e.g. Pravec & Harris 2000); Most asteroids with satellites (*binary asteroids*) in this size range have rotation periods just below the spin barrier, suggesting they formed after rotational disruption (e.g. Pravec et al. 2006); And asteroids were observed while shedding mass and breaking apart (e.g. Jewitt et al. 2014; Drahus et al. 2015) with no collisional scenario to explain it.

By definition, the rubble pile structure dictates a bulk density that is smaller than the material density from which it is made. A good example is 25143 Itokawa, that was measured to have a density of 2.0 gr cm$^{-3}$ by the Japanese spacecraft Hayabusa (Sànchez & Scheeres 2014 based on mass measurement from Fujiwara et al. 2006), while it is composed of LL ordinary chondrites that have a density of $\sim3.3$ gr cm$^{-3}$ (Britt & Consolmagno 2003). Indeed, from Hayabusa's images, Itokawa appears to be constructed of multiple rocks, tens of meters in size, and is not monolithic. Another silicate asteroid, 433 Eros, was found to have a higher density of $2.67\pm0.03$ gr cm$^{-3}$ by the spacecraft NEAR-Shoemaker (Veverka et al. 2000). Larger than Itokawa (mean diameter of 16,840 m compared to $\sim330$ m), Eros is covered by a thin regolith, thus its internal structure is masked. Applying Itokawa's and Eros' density values to calculate the critical spin period before disruption of a rubble pile sphere in a zero-tensile model (Eq. 16 in Pravec & Harris 2007), results with 2.3 h and 2.0 h, respectively, in agreement with the spin barrier at 2.2 h. Other published density values of similar size asteroids are mainly constrained by the orbit of a companion satellite. However, the density uncertainty is high mainly due to limited knowledge of the size and shape of the main body and the satellite's orbital parameters (Carry 2012). Thus the density distribution of rubble piles remains uncertain.

In addition to self-gravity, studies of granular physics have demonstrated that van der Waals cohesive forces between small size grains trapped in a matrix around larger boulders can enlarge the asteroid internal strength and act as a mechanism resisting rotational disruption (e.g. Goldreich & Sari 2009, Scheeres et al. 2010, Sànchez & Scheeres 2014, Scheeres 2015). Indeed, Apollo mission's samples allowed measuring cohesion of lunar regolith (Mitchell et al. 1974), and constraints were placed on the cohesion of Martian regolith by trenches and scuffs created by wheels of rovers (Sullivan et al. 2011). Asteroid cohesion was constrained by grain size of *in situ* measured Itokawa (Sànchez & Scheeres 2014) and by a few fast rotating and intact asteroids (e.g. Rozitis et al. 2014, Polishook et al. 2016). However, it is unclear if these asteroids are representative of the broad class of rubble pile bodies.

The internal strength of a rubble piles must play a larger role in stabilizing asteroids as they spin up. The thermal YORP effect (Rubincam 2000) can spin up km-sized asteroids efficiently in a $10^6$-$10^7$ years by imposing torque from uneven reflected and re-emitted light from an asymmetric shape. It can also align the spin axis of these asteroids to their orbital plane (Vokrouhlický et al. 2003). The effect of YORP was identified on specific asteroids (e.g. Taylor et al. 2007), on the spin axis direction of family members (Slivan 2002) and on the spin distribution of km-sized asteroids (e.g. Pravec et al. 2008, Polishook & Brosch 2009). As it can accelerate rubble piles to beyond the spin-barrier at 2.2 h, the YORP effect is held responsible for the rotational-disruption of km-sized asteroids and the formation of binary asteroids and separated pairs (e.g. Scheeres 2007, Walsh et al. 2008).

Asteroid pairs are those found to share similar orbits but not positions. However, backward integrations show that they were once within each other's Hill sphere (Vokrouhlický & Nesvorný 2008). Lists of almost 200 pairs were found using such dynamical criteria applied on all known asteroids (Pravec & Vokrouhlický 2009; Rozek et al. 2011). Each pair had a single progenitor, with a rubble-pile structured body, that split in the last couple of $\sim 10^6$ years due to the rotational-disruption mechanism (Pravec et al. 2010). Broadband photometry (Moskovitz 2012) and spectroscopy (e.g. Duddy et al. 2013, Polishook et al. 2014, Wolters et al. 2014) of asteroid pairs have shown identical spectral signature between the members of the same pair, supporting their shared origin. In addition, some of the pairs present fresh, non-weathered, reflectance spectra (Polishook et al. 2014), indicating a recent and a significant disruption. Because no spectral variation was detected on young asteroid pairs, Polishook et al. (2014b) concluded that dust covers the primary body homogeneously following a rotational-fission event. Asteroid pairs were found to belong to different taxonomies (e.g. Moskovitz 2012), suggesting that rotational fission is not a function of the asteroid's composition but rather of its structure. Linking the pairs to the YORP effect as the mechanism that spun-up their progenitors is possible when their spin axes are aligned with their orbital planes (Polishook 2014a). About dozen of small clusters, including 3-5 bodies each, were found to be consistent with the rotational fission formation process (Pravec et al. 2018), showing that the rotational disruption might be a complex process, and relevant to the formation of some binary asteroids.

The unique history of asteroid pairs makes them excellent natural laboratories to study asteroid interiors and strength, constraining parameters such as density and cohesion. Here we consider the criteria for asteroid break-up using shape models of asteroid pairs that allow us to calculate the local acceleration magnitude and direction and to construct a map of slopes on a body's surface.

**Methods**

*Observations and data reduction*

The observations were conducted at the Wise Observatory in Israel (Observatory code 097) during multiple apparitions and aspect angles, ranging on over a decade of observations (2007-2019) to improve the constraints on the resulted models of the inversion technique. All observations were performed using the three telescopes of the Wise Observatory: the 1-m Ritchey-Chrétien telescope; the 0.71-m and the 0.46-m Centurion telescopes (Brosch et al. 2008, 2015, respectively). Most observations were performed without filters ("Clear"), or with wide, Luminance filter, though some were done using Cousins *R*. To minimize the point spread function for the moving targets (at a seeing value of 2 to 4 arcsec), exposure times did not exceed 240 seconds, all with an auto-guider.

The images were reduced in a standard way using bias and dark subtraction, and were divided by a normalized flat field image. *IRAF*'s *phot* function was used for the photometry. Apertures with four-pixel radii were chosen to minimize photometric errors. The mean sky value was measured using an annulus of 10 pixels wide and inner radius of 10 pixels around the source's center. The photometric values were calibrated to a differential magnitude level using tens to hundreds local comparison stars measured on every image using the same method as for the asteroid. The brightness of the comparison stars remained constant to ±0.02 mag. The asteroid data were corrected for light-travel time and the magnitudes were reduced to one AU distance from the Sun and the Earth, following the standard procedure (Bowell et al. 1989). Refer to Polishook and Brosch (2009) and Polishook (2014a) for detailed description of the photometric procedures of observation, reduction and calibration.

*Spin Analysis*

The calibrated data per asteroid and per apparition were folded by the known spin periods (Marchis et al. 2008, Pravec et al. 2010, Polishook 2011, 2014b) in order to identify outliers, eclipse events, and other non-periodic features. These measurements were excluded from the dataset before the application of the lightcurve inversion technique, in order to smooth the lightcurves and derive better fit models. Since most of the photometry from different nights was not calibrated on an absolute scale, we fitted a magnitude constant per observing night. The folded lightcurves at each apparition are presented in figures A1-A11 of the appendix.

The lightcurve inversion technique (Kaasalainen and Torppa 2001, Kaasalainen et al. 2001) was implemented as a software package by Durech et al. (2010) to fit multiple solutions of sidereal period, spin axis direction and shape model that are constrained together to the photometric data. This is achieved by measuring the variability of the asteroids' brightness at numerous times, aspect angles and geometric conditions. Spin axes and shape models were derived for hundreds of asteroids using this algorithm and software, and are considered as good representations of the real shape of the asteroids (see many examples on the Database of Asteroid Models from Inversion Techniques (DAMIT) website: http://astro.troja.mff.cuni.cz/projects/asteroids3D).

The algorithm first matches a wide range (0.4 h) of sidereal periods to the data around the already known synodic period, and finds the best sidereal period with the minimal $\chi^2$. We run the algorithm in several iterations, each time focusing on the previously found best period while decreasing the search range and the sampling interval. We find the optimal solution which has $\chi^2_{min}$. We define an acceptable threshold of $\Delta\chi^2$, calculated from the confidence level $\sigma$, and the square root of the variance normalized by the number of degrees of freedom $\upsilon$,

$$\Delta\chi^2 = \sigma \frac{\sqrt{2\upsilon}}{\upsilon} \quad,$$

since a $\chi^2$ distribution with $\upsilon$ degrees of freedom has a variance of $2\upsilon$ (Press et al., 2007). We chose this approach to determine the statistical uncertainty, following Vokrouhlicky et al. (2011, 2017) that considered the problem of photometric uncertainties that may not obey Gaussian statistics and the problem of systematic and modeling errors dominating over the random ones. In this way, the uncertainty on the sidereal period we derived reduces with increasing number of measurements; for our sample the threshold $\Delta\chi^2$ is 5% to 15% of $\chi^2_{min}$. For all asteroids we used a confidence level of $\sigma = 3$. We regard all solutions with $\chi^2 < \chi^2_{min} + \Delta\chi^2$ as acceptable. We report only cases where a unique solution was found for the sidereal period.

For each asteroid, we then run a grid search on the spin axis coordinates ($0^o$ to $360^o$, -$90^o$ to $90^o$ in steps of $5^o$) to find the best fitting photometric curve, with the same acceptance threshold as before. With this method we derived a range of statistically possible spin axes per asteroid. We present here only asteroids for which the spin axes distribution was constrained to better than a hemisphere (Fig. 1). Even though some solutions better fit the photometric data than others, we chose a conservative approach and present the 5th and the 95th percentile of the entire distribution as the lower and upper limits for the sidereal period and spin axis coordinates.

Using the derived spin axes we calculate the range of the obliquity $\phi$ for each asteroid:

$$cos\phi = \sin\beta_{pole} \sin\beta_{orb} + \cos\beta_{pole} \cos\beta_{orb} \cos(\lambda_{pole} - \lambda_{orb})$$

where, $\lambda_{pole}$ and $\beta_{pole}$ are the spin axis longitude and latitude, respectively, and $\lambda_{orb}$ and $\beta_{orb}$ are the longitude and latitude of the normal to the orbit of the asteroid, defined as $\lambda_{orb} = \Pi - 90^o$ and $\beta_{orb} = 90^o - i$ ($\Pi$ is the longitude of ascending node and $i$ is the orbit's inclination). For each asteroid we present the 5th and the 95th percentile of the obliquity values for all valid spin axis solutions.

*Shape construction*

Shape models are constructed from photometric measurements using the lightcurve inversion code mentioned above. In order to quantify the elongation and flattening of the shape we fit a triaxial ellipsoid to the model denoting the three axes with $A$, $B$ and $C$ where $A \geq B \geq C$. Since none of the observed asteroids presents a complex lightcurve that can be modeled with non-principal axis rotation (tumbling rotation; Pravec et al. 2005), we assume that all of the objects rotate around their maximum moment of inertia that is approximately axis $C$. We calculate the three principal moment of inertia, $I_x$, $I_y$, $I_z$ (assuming uniform density) and the angles between these axes to the spin axis. We reject shape models on the basis of non-physical solution, if they violate at least one of the following. We separate cases where one inertia components dominates, from cases where two or all three components are similar:

- If $I_z > I_y, I_x$, and the angle between the spin axis to $I_z$ is smaller than 30 degrees.
- If $I_z \sim I_y > I_x$, and the angle between the spin axis to $I_x$ is larger than 60 degrees (that is the angle with respect to the $I_z - I_y$ plane is again smaller than 30 degrees.
- $I_z \sim I_y \sim I_x$, no restriction is applied.

we use a criterion of $30^o$ following Hanus et al. 2011. We consider inertia components similar when the difference between them is smaller than 15%.

The spin axis solutions that passed these criteria are presented in Fig. 2. We list the number of valid shape models for each asteroid with the 5th and the 95th percentile of the ratios $A/B$ and $B/C$. In Figs. 3 and 4 we present the shape models with the lowest $\chi^2$ only, and note that other shape models within the determined uncertainty for each asteroid are similar to those presented.

*Slopes calculation*

For every valid shape model, we map the gravitational and rotational forces on each facet. This allows us to construct a map of the slopes on the shape models' surfaces.

We calculate the gravity vectors $\vec{f_g}$ by integrating a constant density model, and relying on Gauss's theorem to replace the volume by a surface integral:

$$\vec{f_g} = -\int \frac{G\rho\vec{r}}{|r|^3} dV = \int \vec{F} \cdot d\vec{a}$$

where $\vec{F} = G\rho \frac{\vec{r} \otimes \vec{r}}{|r|^3}$ is the tensor that its divergence equals the gravity force

$$\vec{\nabla} \cdot \vec{F} = -\frac{G\rho\vec{r}}{|r|^3} \ ,$$

$\vec{r}$ is a facet vertex position, $d\vec{a}$ is directed along the outwards facet normal with a magnitude of its area, $\rho$ is the bulk density and $G$ is the gravitational constant and $\otimes$ represents the outer product operator. The gravity vectors at the three vertices of a facet, are averaged to derive the gravity vector of the facet. The density is assumed to be 2 gr cm$^{-3}$, (similar to the density of the rubble-pile structured asteroid 25143 Itokawa), or a density of 2.7 gr cm$^{-3}$ (average density of S-type asteroids and the density of 433 Eros; Carry 2012).

The centrifugal vector for every facet is the average of the centrifugal forces of the facet's vertices, $\vec{f_\omega}$. Using the spin rate $\omega$,

$$\vec{f_\omega} = \omega^2 \vec{r}$$

The acceleration term $\vec{f_a}$ is a simple addition of the gravity and centrifugal force:

$$\vec{f_a} = \vec{f_g} + \vec{f_\omega}$$

The local slope of a facet $\theta$ is defined to be the angle between the inwards surface normal of a facet and the local acceleration term:

$$\cos\theta = -\frac{d\vec{a} \cdot \vec{f_a}}{|d\vec{a}| \left|\vec{f_a}\right|}$$

An example for the acceleration vectors and the slopes calculation is presented at Fig. 5, where the acceleration vectors of a slow- and fast-rotating bodies are compared. The fast rotating case shows the direction of the local acceleration vectors changes relative to the slow-rotating case, increasing the local slope. For the slow-rotating case, the slope map and associated histogram show that vast majority of the slopes are below the angle of repose.

*Disruption model*

For each asteroid in the set, we determine the rotation rate at which frictional failure occurs. Our criterion is reached when a region on the shape model of an area half the surface area of the secondary member has a slope larger than the angle of repose. This criterion was chosen following a simplifying assumption, in which the secondary is considered to be an ejected fragment which originally had half its surface area in contact with the primary. The angle of repose is set to 40$^{\text{o}}$ - the angle of friction of lunar regolith (Mitchell et al. 1974). This is a conservative choice as it represents the upper limit of published values (35$^{\text{o}}$ for the mean of geological materials, Hirabayashi & Scheeres 2015; 30$^{\text{o}}$ to 37$^{\text{o}}$ for Mars regolith, Sullivan et al. 2011; 30$^{\text{o}}$ - 40$^{\text{o}}$ for list of *in situ* measured asteroids including Eros and Itokawa, Richardson & Bowling 2014; 40$^{\text{o}}$ is the upper limit for 99% of 433 Eros' surface, Zuber et al. 2000).

We iteratively searched for the minimum rotation period in which a contiguous region on the primary has a slope exceeding the angle of repose. The regions we found indicate potential locations of the detachment of the secondary. We limited the regions to occupy a single hemisphere in order to avoid the non-physical case of the detachment area encircling the primary. The potential detachment regions usually occur near the two ends of an elongated asteroid. The choice of half the area for the criterion is conservative; smaller areas would lead to a higher critical rotation period. This is equivalent to an increased size of the secondary, studied below. We also note that our simplifying assumption of two components laying one on top the other is theoretic − while it is physically reasonable for pairs with small secondaries (e.g. 5026, 7343), pairs with large secondaries (e.g. 4905, 88604) have probably more complex structures that includes part of the secondary buried within the primary object (such as Itokawa's "head"), thus the disruption process is more complicated. However, even in this case, the local slope at each of the contact facets should be higher than the angle of repose to allow disruption, in agreement with the criterion of our model. Therefore, we think our model can derive the disruption rate at good order of magnitude.

We provide the resulting rotation periods at break-up per asteroid, and present the shape models with a map of slopes at disruption, with their secondary members at the measured relative sizes, for comparison (Fig. 6).

*Sensitivity to shape model*

We seek to evaluate the extent to which our disruption model is sensitive to the shape solution. We perform several tests: first, we test the contribution of the shape's elongation and flattening to the critical period; then we examine how a convex model with some roughness or irregularities (the shape models we derive by the inversion algorithm) affects the critical rotation period compared to a smooth triaxial ellipsoid, that is usually available from standard photometric observations; Finally, we test the effect of the convex hulls we obtained, compared to concave shape models that often occur on real asteroids. In addition, we compare the derived shapes of pairs to those of non-pairs.

Elongation and flattening dependence:

Since elongated objects are less stable, we expect a correlation between the critical rotation period and elongation (i.e. the physical axes ratio $A/B$). Harris (1996), estimated this correlation as $P_{\text{crit}} \propto \sqrt{A/B}$. Therefore, we run our analysis code on all the derived shape models, with all other parameters normalized to a single value (effective radius of $R$=3 km, a hypothetical secondary with a radius of 1.2 km, equivalent to magnitude difference $dH$=2, density $\rho$=2 g cm$^{-3}$; the approximate averages of the real asteroids in our sample).

Fig. 7 displays the physical axes ratios vs. one another ($A/B$ vs. $B/C$, when $A \geq B \geq C$), with the markers' colors representing the critical period. The correlation is clear with a linear fit of $P_{\text{crit}} \sim (0.62\pm0.02) \ A/B + (0.39\pm0.02) \ B/C$ + constant, where the constant is a function of the other parameters (density, secondary size, etc.). Therefore, a spherical shape will have a minimal critical rotation period (the most stable shape), a pie/pancake shape is less stable, following by a cigar-shape body. The most unstable shapes, breaking at slower rotation rates are cases where both $A/B$ and $B/C$ are large. Future study may test if such bodies are indeed less common.

Convex models vs. simple triaxial models:

In order to isolate the effect of the model roughness, we used the three axes of each derived shape, $A$, $B$ and $C$, but considered a smooth triaxial ellipsoids with these axes. We normalized the pairs parameters to a single value as previously. We found that most (~95%) of the triaxial shape models disrupt at faster rotation compared to the inversion-based shapes, though the difference is subtle, with a median of 0.08 h and the 5th and the 95th percentile ranges between 0 to 0.17 h (Fig. 8). We associate this difference with the higher slopes of the convex models, making them less stable.

Convex models vs. concave models:

We fit a convex model to our observations, which in some cases may represent a shell surrounding an actually concave object (for example, Eros and Itokawa have significant concavities, while Bennu and Ryugu do not). In order to check the effect of this assumption on our disruption model, we run the model on a random selection of twenty asteroids from the DAMIT database, for which both convex and concave models were published (Kaasalainen et al. 2002, Torppa et al. 2003, Carry et al. 2010, Durech et al. 2011, Hanus et al. 2013, 2016, 2017, Franco & Pilcher 2015, Viikinkoski et al. 2017). While keeping the shape parameters, we applied the size parameters of all asteroids in our sample to each of the 20 shapes. We found that the convex versions of these models are slightly more stable than the concave ones, with a median difference of 0.03 h longer rotation period in the concave models and a 5th and a 95th percentile ranges between -0.085 to 0.215 h (Fig. 9).

Models of observed pairs vs. non-pairs:

In order to examine if asteroid pairs have shapes that are more prone to disruption, we compared the critical rotation period of the asteroid pairs' shape models to the convex shape models of other asteroids from the literature (the DAMIT website), with all other parameters normalized to a single value, as described above. The two distributions are similar (Fig. 10), suggesting that asteroid pairs shapes are representative of non-pairs.

**Results**

*Observed sample*

Our current sample includes observations of eleven primary members of asteroid pairs that were selected for this work from the list of known pairs (e.g. Vokrouhlický & Nesvorny 2008, Pravec & Vokrouhlický 2009) because they were frequently available to the Wise Observatory's telescopes and unique solutions for them were derived. Each asteroid was observed during three to seven apparitions, on ~10-30 nights, resulting in hundreds to thousands of measurements. This allows us to derive shape models that are statistically and physically valid. Observations used in this study were acquired from 2007 up until 2019, all performed at the Wise Observatory by DP. The shape models of five of the asteroids discussed here where also published by Polishook (2014a) but with larger uncertainties. The shape models of some of the asteroids presented here were also presented in Vokrouhlický et al. (2011, 2017), and Pravec et al. (2019) with data collected from other stations; our results are in agreement with their spin axes values. The observational circumstances are summarized in the appendix's Table A1. Some of the photometric observations were previously published by Pravec et al. (2010), Polishook (2011), Polishook et al. (2011), Vokrouhlický et al. (2011) and Polishook (2014a, 2014b). Some of this previously published data is summarized in Table 1 of Polishook (2014a) and not in the current paper, though it appears in the lightcurves of Fig. A1-A11.

The observed asteroids represent the large end of asteroid pairs with diameter range of ~3-10 km, and a wide range of diameter ratio (secondary over primary) of 0.1-0.6. Most are S-type asteroids (thus their material density is comparable to S-complex asteroids such as Eros and Itokawa), though three asteroids of the sample present reflectance spectra of a Ch-type, E-type, and C or X-complexes. The orbital and physical parameters of the sampled asteroids are detailed in Table 1. The orbital data was derived from the Minor Planet Center, and the taxonomy from the literature (Polishook et al. 2014, Duddy et al. 2012). The diameters were estimated from the absolute magnitude assuming an albedo value of 0.197 for S-complex asteroids (Pravec et al. 2012), 0.36 for V-type, 0.43 for E-type, 0.05–0.15 for C/X-complex and 0.058 for the Ch-type asteroid (Mainzer et al. 2011). We compared these diameters to the values collected by the Wide-field Infrared Survey Explorer (WISE; Masiero et al. 2011), after recalculating them with updated absolute magnitude H. The two asteroids in our sample (2110 and 4905) that were measured by WISE with 3 or 4 channels, i.e. those with secure results, are 10% and 5% off the diameters we calculated, suggesting our approach is correct.

Table 1: Orbital and physical parameters of the observed asteroid pairs

| Asteroid | $a$ [AU] | $e$ | $i$ [deg] | $H$ [mag] | $D_1$ [km] | Taxonomy | Secondary | $dH$ [mag] | $D_2$ [km] | Age [kyrs] |
|----------|----------|-----|-----------|-----------|------------|----------|-----------|------------|------------|------------|
| 2110 | 2.198 | 0.177 | 1.130 | 13.4 | 6.3 | S | 44612 | 2.1 | 2.4 | > 1,600 |
| 3749 | 2.237 | 0.109 | 5.382 | 13.3 | 6.6 | Sq | 312497 | 4.4 | 0.9 | $280^{+45}_{-25}$ |
| 4905 | 2.601 | 0.169 | 12.426 | 12.2 | 10.9 | Sw | 7813 | 0.9 | 7.2 | > 1,650 |
| 5026 | 2.378 | 0.244 | 4.288 | 13.9 | 9.2 | Ch | 2005WW113 | 3.9 | 1.5 | 18 ± 1 |
| 6070 | 2.387 | 0.211 | 3.130 | 13.8 | 5.2 | Sq | 54827 | 1.6 | 2.5 | 17 ± 0.5 |
| 7343 | 2.193 | 0.139 | 3.959 | 14.0 | 4.7 | S | 154634 | 2.8 | 1.3 | > 800 |
| 10484 | 2.320 | 0.079 | 5.733 | 13.9 | 3.7 | V | 44645 | 1.1 | 2.2 | $310^{+210}_{-50}$ |
| 16815 | 2.559 | 0.022 | 11.052 | 12.7 | 10 - 17 | C/X | 2011GD83 | 4.3 | 1.4 - 2.4 | $95^{+40}_{-20}$ |
| 25884 | 1.954 | 0.081 | 21.564 | 14.7 | 4.6 | Xe/E | 48527 | 1.4 | 2.4 | $420^{+200}_{-100}$ |
| 42946 | 2.568 | 0.071 | 4.689 | 13.8 | 5.2 | Srw | 165548 | 1.9 | 2.2 | $600^{+580}_{-150}$ |
| 88604 | 2.668 | 0.107 | 11.757 | 13.4 | 6.3 | S/Sq | 60546 | 1.3 | 3.4 | > 1,000 |

*Sidereal rotation period and spin axis*

The lightcurve inversion code derives statistically valid results with small uncertainty only if observations are collected on at least ~3-4 apparitions with a total of ~700 measurements spread around these apparitions. The quality of the results depends on the S/N of the observations and the variation in the viewing geometry, thus the number of observations above may be insufficient if the asteroid is sampled with a low S/N or over a narrow range of geometries. The algorithm found a sidereal period for each asteroid with small uncertainties (median ~0.01 sec) at a confidence level of $\sigma = 3$. We present the $\chi^2$ values for all spin axis solutions in ecliptic longitude–latitude coordinates for each asteroid in our sample (Fig. 1), and a similar figure with the spin axes solutions only for models that have feasible physical parameters as defined previously (Fig. 2). Studying the derived spin axes distribution, we find that eight asteroids show retrograde rotation, and three show prograde rotation. The spin axis longitude of five of the asteroids are mostly degenerate, because the pole is almost orthogonal to our aspect angle, or due to small number of viewing geometries. For the other six asteroids, the longitude is constrained to two mirror solutions, and we give them both in Table 2.

The median obliquity of all the asteroids is closer to 0° or 180° than to 90° although in some cases (3749, 7343, 10484, 16815, 42946, 88604) the uncertainty on the spin axis is too large to decide if it is closer to the pole or the equator. We note that asteroids with obliquities close to 0° or 180° are consistent with independent observations and with the theory that describes the YORP effect as the physical mechanism responsible to align the asteroids spin axes to their orbit (Vokrouhlicky et al. 2003). Indeed, Slivan (2002) discovered that the spin axes of the Koronis asteroid family are trapped in spin-orbital resonances, one at positive and the other at negative latitude. Hanus et al. (2011, 2013) have found that the obliquities of main belt asteroids are not evenly distributed, with higher fraction close to 0° and 180°. Since the YORP effect can also lead to the ultimate disruption of rubble piles forming asteroid pairs, the notion is that asteroid pairs with obliquities at approximately 0° or 180° maintained the obliquities of their progenitors after disintegration. Polishook (2014a) and Vokrouhlicky et al. (2017) showed that the primary members of the pairs 2110-44612 and 6070-54827, respectively, did not only preserve the low obliquity of their progenitors but also kept rotating in the same sense as their secondaries (a retrograde rotation for both pairs), suggesting a gentle breakup.

Table 2: Derived values from the lightcurve inversion technique from all valid models per asteroid: median of the sidereal period; ranges based on the 5th and the 95th percentile of the spin axis longitude and latitude, obliquity, triaxial shape ratios parameters (*A/B* and *B/C*) and the number of valid models.

| Asteroid | Sidereal Period [h] | Spin axis longitude [deg] | Spin axis latitude [deg] | Obliquity [deg] | *A/B* | *B/C* | Number of models |
|---|---|---|---|---|---|---|---|
| 2110 | 3.3447300±0.000002 | 0 – 360 | -85 - -65 | 155 – 175 | 1.12 - 1.33 | 1.04 - 1.44 | 42 |
| 3749 | 2.8049170±0.000001 | 0 – 360 | -80 - -45 | 135 - 170 | 1.01 - 1.09 | 1.04 - 2.00 | 40 |
| 4905 | 6.04483±0.00002 | 20 – 85 | -85 - -55 | 140 – 165 | 1.18 - 1.39 | 1.06 - 1.20 | 41 |
| | | 150 – 310 | -85 - -40 | 140 - 165 | 1.24 - 1.45 | 1.07 - 1.23 | |
| 5026 | 4.4240860±0.000004 | 0 – 25 | 60 - 80 | 15 - 35 | 1.69 - 1.76 | 1.07 - 1.21 | 13 |
| | | 190 – 200 | 45 - 65 | 20 – 40 | 1.49 - 1.59 | 1.08 - 1.12 | |
| 6070 | 4.2737140±0.000002 | 85 - 175 | -85 - -55 | 150 - 175 | 1.14 - 1.28 | 1.08 - 1.56 | 18 |
| | | 250 - 345 | -85 - -60 | 145 – 170 | 1.16 - 1.27 | 1.01 - 1.55 | |
| 7343 | 3.75494±0.00001 | 15 - 80 | 35 - 75 | 15 - 55 | 1.06 - 1.17 | 1.01 - 1.43 | 52 |
| | | 220 – 265 | 25 - 70 | 15 – 60 | 1.12 - 1.21 | 1.01 - 1.15 | |
| 10484 | 5.5100220±0.000008 | 0 – 360 | -80 - -40 | 130 - 170 | 1.02 - 1.11 | 1.02 - 1.35 | 150 |
| 16815 | 2.9177120±0.000007 | 0 – 360 | -75 - -15 | 105 – 165 | 1.04 - 1.17 | 1.02 - 1.49 | 230 |
| 25884 | 4.9172480±0.000005 | 70 – 95 | -55 - -45 | 145 - 160 | 1.34 - 1.40 | 1.36 - 1.46 | 21 |
| | | 145 - 185 | -65 - -40 | 140 – 175 | 1.34 - 1.42 | 1.29 - 1.62 | |
| 42946 | 3.4081230±0.000002 | 0 - 360 | 25 – 75 | 15 - 65 | 1.10 - 1.29 | 1.03 - 1.82 | 117 |
| 88604 | 7.1723980±0.000004 | 35 – 70 | -65 - -20 | 120 – 165 | 1.48 - 1.64 | 1.12 - 1.41 | 52 |
| | | 245 - 310 | -70 - -40 | 125 - 160 | 1.48 - 1.69 | 1.05 - 1.46 | |

*Shape models*

Our algorithm derived multiple shape models for each asteroid that are all compatible with the observations within the uncertainty. However, in some cases the differences between the models are subtle, in one or more dimensions. For example, all of the 13 possible shape models of asteroid 5026 are similar to one another. In other cases, the flattening of the models ($B/C$) are less constrained resulting with some variation in the axis. With scanning interval of five degrees at the longitude-latitude plane, the total number of possible models was $72 \times 37 \times 11 = 29,304$. From these, the inversion algorithm derived 776 acceptable models for the entire asteroid sample. The 5th and the 95th percentile of the triaxial ratios of the valid models of each asteroid, $A/B$ and $B/C$, are summarized in Table 2.

Among our sample, seven asteroids have elongated shapes (median $A/B$ of 1.37), and four others (3749, 7343, 10484, 16815) are roughly spherical (median $A/B$ of 1.09) with shape models resembling the "top-shaped" bodies such as (66391) 1999KW4 (Harris et al. 2009). This is not necessarily representative of the real distribution of asteroid shapes. There is a bias inherent in the lightcurves of spherical asteroids that show little variations and hence provide fewer constraints on the spin solution. However, finding both spherical and elongated shapes suggests that rotational-disruption resulting from spin-up does not preferentially result in a top-shaped model, as stated by other studies (*e.g.* Statler 2015). Both top-shaped and elongated bodies can lose mass and form asteroid pairs.

*Critical rotation periods*

Our 776 spun-up models reach the disruption criteria at a range of values from 2.6 to 3.5 h. Table 3 presents the median value of the critical rotation period per asteroid, with the median for all asteroids at 2.8 h. When the results are median averaged for each asteroid, the critical rotation period ranges between 2.6 to 3.0 h only, when 5026 Martes is excluded. This asteroid is unique in its very elongated shape ($A/B$ ranges between 1.69 to 1.76 or 1.49 to 1.59) and in its small secondary ($D_2/D_1 = 0.15$), thus, a small area with high slope is required to reach the disruption criterion. Another two asteroids, 3749 Balam and 16815 1997UA9, also have small secondaries ($D_2/D_1$ of 0.13, and 0.14, respectively) but their shapes are rounded ($A/B$=1.01 to 1.09, and 1.04 to 1.17, respectively).

While the disruption rotation period we derived is comparable to the classic spin barrier at 2.2 h (Pravec & Harris 2000), it is slower by ~30%, implying additional effects and/or parameter adjustment are required in our disruption model. The shape difference among the possible models can account for only a limited change of the critical rotation period, as shown above. Below we discuss some possibilities that could account for this mismatch.

**Discussion**

In order to explain our result of slower critical rotation period than the commonly quoted 2.2 h, we suggest several alternative scenarios, noting that combination of these is also possible. The scenarios detailed below invoke an ejected body that was originally larger than the current secondary component, a higher value of density, a higher angle of repose and the presence of cohesion. These potential explanations are required if one assumes that asteroid pairs actually break up at the spin barrier of 2.2 h as determined for non-pairs. This assumption relies on pairs being representative of asteroids in general in their physical properties (e.g. size, structure, composition, cohesion) and history (e.g. thermal spin-up), and they are distinguished only in the time of the disruption occurring relatively recently ($\lesssim 10^6$ years). It is possible, however, that asteroid pairs are actually weaker than the standard asteroid (in the same size range), and thus disrupt at a spin barrier that is longer than 2.2 h.

Table 3: Derived critical rotation period, required cohesion.

| Asteroid | $\rho = 2.0$, measured $D_2/D_1$ | | $\rho = 2.0$, maximal $D_2/D_1$ | | $\rho = 2.7$, measured $D_2/D_1$ | | $\rho = 2.7$, maximal $D_2/D_1$ | |
|---|---|---|---|---|---|---|---|---|
| | Critical Rotation Period [h] | Cohesion [Pa] | Critical Rotation Period [h] | Cohesion [Pa] | Critical Rotation Period [h] | Cohesion [Pa] | Critical Rotation Period [h] | Cohesion [Pa] |
| 2110 | 2.71 - 3.02 | 430 – 890 | 2.57 - 2.80 | 480 - 860 | 2.33 - 2.60 | 230 - 700 | 2.21 - 2.40 | 40 - 480 |
| 3749 | 2.76 - 3.21 | 130 – 390 | 2.53 - 2.95 | 450 - 1450 | 2.38 - 2.76 | 70 - 350 | 2.17 - 2.54 | 20 - 930 |
| 4905 | 2.66 - 2.78 | 2150 – 3410 | 2.66 - 2.78 | 2150 - 3410 | 2.29 - 2.40 | 740 - 1790 | 2.29 - 2.40 | 740 - 1790 |
| 5026 | 3.26 - 3.50 | 700 – 900 | 2.66 - 2.86 | 1760 - 3160 | 2.81 - 3.01 | 670 - 920 | 2.29 - 2.46 | 610 - 2090 |
| 6070 | 2.67 - 2.85 | 420 – 710 | 2.60 - 2.75 | 420 - 680 | 2.30 - 2.45 | 180 - 440 | 2.23 - 2.37 | 70 - 340 |
| 7343 | 2.79 - 2.98 | 140 – 320 | 2.52 - 2.72 | 230 - 460 | 2.40 - 2.56 | 80 - 240 | 2.17 - 2.34 | 20 - 220 |
| 10484 | 2.57 - 2.73 | 150 – 270 | 2.57 - 2.73 | 150 - 270 | 2.21 - 2.35 | 10 - 120 | 2.21 - 2.35 | 10 - 120 |
| 16815 | 2.85 - 3.14 | 570 – 1380 | 2.57 - 2.77 | 2070 - 4310 | 2.45 - 2.70 | 450 - 1170 | 2.21 - 2.39 | 150 - 2270 |
| 25884 | 2.82 - 2.95 | 490 – 700 | 2.72 - 2.85 | 490 - 710 | 2.42 - 2.53 | 340 - 480 | 2.34 - 2.45 | 260 - 430 |
| 42946 | 2.70 - 3.06 | 330 – 780 | 2.62 - 2.83 | 340 - 890 | 2.32 - 2.63 | 160 - 650 | 2.26 - 2.43 | 80 – 570 |
| 88604 | 2.72 - 2.93 | 1020 – 1600 | 2.68 - 2.85 | 950 - 1550 | 2.34 - 2.52 | 520 - 1220 | 2.31 - 2.45 | 380 – 940 |

*Larger original ejected body*

Since in our model, disruption occurs when the area with high slopes is larger than half the surface of the secondary, an original secondary component that is larger in size than at present would require a larger area with high slopes on the primary and hence a faster critical spin rate. The idea that the secondary is larger at breakup was suggested by Jacobson and Scheeres (2011) that studied the effect of tidal forces acting on the ejected component in the vicinity of the main body. Using numerical calculations, they showed that ~40% of the rotationally disrupted asteroids suffer additional disruptions, and of these cases, some components are separated thus forming asteroid pairs.

In order to study the effect of a larger secondary on the critical rotation period we run our analysis code using different values of the absolute magnitude difference $dH$. This can be translated to diameter ratio ($D_2/D_1$) and mass ratio ($M_2/M_1$) using $D_2/D_1 = 10^{-0.2dH}$ and $M_2/M_1 = (D_2/D_1)^3$, assuming the two pair members have the same albedo and density values. We use the measured $dH$ value and compared the resulted critical rotation period to a scenario with a maximal $dH$ value allowed for asteroid pairs. A maximum mass ratio of $M_2/M_1 \sim 0.2$ for asteroid pairs (equivalent to $D_2/D_1 \sim 0.6$) was determined theoretically by Scheeres (2007) that calculated the amount of free energy in the system that allows the secondary to escape. This threshold was confirmed observationally by spin measurements of asteroid pairs (Pravec et al. 2010).

Increasing $D_2/D_1$ is most effective for asteroids with low $D_2/D_1$ ratio such as 5026 Martes that has a $D_2/D_1 = 0.16$, and thus its critical rotation period decreases by 0.6 h relative to the maximal value. However, since most of the asteroid pairs already have secondary components that are almost at the maximal threshold, changing this parameter has usually little effect, on average ~0.1 h, on the critical spin period. Figure 12 presents the decrease in critical rotation period for every asteroid and model we derived. Therefore, we conclude that while originally larger secondaries are possible, this effect alone is insufficient to modify the critical rotation period to 2.2 h, and other reasons are required in order to reach the spin barrier.

Interestingly, another pair in our sample with small secondary ($D_2/D_1 \sim 0.13$) is 3749 Balam with a median critical rotation period of 2.90 h, which is problematic since its current rotation period is faster (~2.805 h). However, this asteroid is unique: it has at least two satellites in addition to the ejected component that defines it as an asteroid pair. Both satellites are larger than the secondary pair ($D_2/D_1 \sim 0.4$ for each of the satellite; Marchis et al. 2008, Polishook et al. 2011). When assuming the two satellites and the secondary pair were all ejected together from 3749 as a single body (summing the masses of the secondary pair and the two satellites results with $M_2/M_1 \sim 0.13$, $D_2/D_1 \sim 0.5$), our model results with a faster critical period with a median of 2.8 h, equivalent to 3749's current rotation period. This supports the idea that indeed the two satellites and the secondary pair of Balam separated at the same time as described by the Jacobson & Scheeres model. This also suggests that the secondary of 5026 is also a small fragment of a larger body that was ejected from the main asteroid. The larger ejected component might not have been identified yet, or, it crashed back on the surface of the primary asteroid, elongating its shape.

In order to examine the effect of the secondary's gravity acting on the primary's surface, we computed the effect of a point mass placed near the facet with the highest slope of the primary, and repeated our procedure above for this perturbed gravity field. Introducing the secondary mass resulted with the critical rotation period increasing by only 1 to 4%, depending on the size of the secondary. Therefore, we conclude that the secondary mass has little effect on the critical rotation period.

*Higher density*

Increasing the assumed density of the asteroids from Itokawa's value of 2.0 g cm$^{-3}$ to Eros' value of 2.7 g cm$^{-3}$, which is also the average density of S-type asteroids (Carry 2012), increases the self-gravity and the critical rotation rate. If indeed disintegrated by rapid spin, the asteroids are expected to be rubble piles and their densities must be lower than ordinary chondrite's density (3.3 g cm$^{-3}$, Carry 2012). We found that the critical rotation period decreased by ~0.4 h to ~2.4 h on average, compared to the parallel scenario with Itokawa-like density. With a maximal $D_2/D_1 \sim 0.6$ and an Eros-like density, the critical period decreases further by ~0.1 h to ~2.3 h, on average, with a standard deviation of 0.05 h. This result is also relevant for the asteroids in our sample that are not S-types (but have similar density values), and specifically to 5026 that supposedly has lower density since it is classified as a Ch-type (their density on average is $1.41 \pm 0.29$ g cm$^{-3}$; Carry 2012). With such a low density it reaches a disruption at 3.99 h and 3.24 h with a maximal $D_2/D_1 \sim 0.6$, which is compatible with the impression that C-type asteroids have, on average, lower limit on their spin rate compared to the S-type asteroids as presented by Chang et al. (2015).

We note that the studied asteroids are an order of magnitude larger than Itokawa (effective diameter 0.320 km), closer in size to Eros (16.3 km) and thus a higher density may be appropriate for the pairs if density increases with size, as Carry (2012) pointed out. One option is that larger asteroids have 'cores' that are more compact and therefore are denser than their 'shells' (e.g. Walsh et al. 2012, Hirabayashi et al. 2015). Further study is needed to constrain the pairs' densities.

*Higher angle of repose*

Increasing the angle of repose required for failure would also results in a faster critical period. Measurements of the angle of friction of Lunar regolith (~40°; Mitchell et al. 1974), Martian sand (30° to 37°; Sullivan et al. 2011) and geological material on the Earth (30° to 40°; Lambe & Witman 1969), are similar to the angle of friction value we used in our study (40°). Even with a maximal value of 90° for the angle of repose, attributed only to wet sand on the Earth and unlikely for asteroid regolith, the critical rotation period we derive is 2.4 h on average, away from the 2.2 h spin limit. Hence, this property is unlikely to account for our results.

*Cohesion*

Significant internal cohesion among the components of the asteroid acts as an additional force to resist the rotational acceleration and thus allow the asteroid to disrupt at a faster rate (e.g. Holsapple 2007, Scheeres et al. 2010, Rozitis et al. 2014). To estimate the amount of cohesion required to set the critical rotation period to 2.2 h, we calculate the additional rotational acceleration added to the system, $\overrightarrow{\Delta f_a}$, at rotation period of 2.2 h, over the value where our disruption criteria is met, $\overrightarrow{f_a}$. Choosing a simple approach, following Hirabayashi & Scheeres (2015), the stress acting on the failure plane is the product of $\overrightarrow{\Delta f_a}$, and the mass of the secondary, divided by the size of the failed area, where for the latter we used half of the secondary surface, $S$, as above for simplicity. We calculate $\overrightarrow{\Delta f_{a_i}}$ per facet $i$, for only the facets that are within the failure area, scaled by the ratio of the facet area by the failure area, $S_i/S$. Integrating over the contributing facets gives the additional pressure on the failed area that is required to maintain the secondary object to the primary asteroid at a rotation of 2.2 h. Since the cohesion acts tangentially to the surface, in computing the minimal required cohesion, $k_{min}$, we use only the tangential component of $\overrightarrow{\Delta f_{a_i}}$ by multiplying it by $\sin \theta_i$, where $\theta_i$ is the slope of the facet:

$$k_{min} = \frac{m_2}{S} \sum_i |\overrightarrow{\Delta f_{a_i}}| \sin \theta_i \, S_i/S \,,$$

where $S = 2\pi R_2^2$, and $R_2$ is the effective radius of the secondary. The derived cohesion values heavily depend on the asteroid's elongation and size of the secondary, and range between ~130 to ~3400 Pa, with the median of ~580 Pa (values detailed in Table 3). When increasing the density from 2.0 (Itokawa-like) to 2.7 gr cm$^{-3}$ (S-type average-like), the required cohesion drops to an average of ~360 Pa with a 5 to 95 percentile range of 10 to 1790 Pa. These cohesion values are comparable to the cohesion of regolith on the Moon (100-1000 Pa; Mitchell et al. 1974), Mars (0-2000 Pa; Sullivan et al. 2011) and other asteroids (e.g. 25143 Itokawa, P/2013 P5, and (29075) 1950 DA with values ranging between 25-100 Pa; Sànchez & Scheeres 2014, Rozitis et al. 2014). Note that all but five of the 776 shape models required some cohesion in order to reach breakup rotation rate of 2.2 h. However, we note that only 1% of same-size known asteroids (0.2 > D > 10 km, not in pairs) spin faster than 2.3 h, and only 5% spin faster than 2.6 h (see Fig. 12). Thus, it may be the case that not all asteroid reach the 2.2 h limit used here.

**Summary**

By mapping the acceleration vectors on constructed shape models of rotationally disrupted asteroids, we were able to calculate local slope values on the asteroid shapes. Compared to the known angle of repose, we derived the critical rotation period at disruption, assuming the area with slope higher than the angle of repose is as large as half the surface of the secondary component. In order to reach critical rotation periods that are similar to the classic rubble pile spin limit of 2.2 h, we investigated the effects of various asteroids' properties including density, secondary size fraction, angle of repose, and cohesion. In addition, we studied the effects of the asteroid shape. We find that:

- Among the five asteroids with small uncertainty on their obliquity, all have obliquity ranges that are closer to $0^o$ or $180^o$ than to $90^o$. This is consistent with the YORP effect as the physical mechanism responsible for the spin-up and the spin alignment with the ecliptic pole and thus supports the notion asteroid pairs were disintegrated by the YORP effect.

- The elongation and flattening of the asteroid's shape can significantly increase the critical rotation period, even by almost an hour relative to a spherical object. Although the critical rotation period is more sensitive to the elongation ($A/B$), the flattening ($B/C$) is also effective and can account for about half the effect.

- The uncertainty of our analysis that comes from the convexity limitation of the models, from the roughness of a model compared to a smooth triaxial body, and from neglecting the secondary mass, is limited to about ~0.2 h.

- Density plays a significant role in determining the critical rotation period. A density of 2.7 gr cm$^{-3}$, the average density of S-type asteroids (and density of 433 Eros), decreases the critical rotation period by ~0.4 h on average, compared to an Itokawa-like density of 2 gr cm$^{-3}$. If the asteroid pairs analyzed here broke up at a spin period similar to the breakup speed usually observed (~2.2 h), then we predict that the density of such km-sized objects is high, more similar to Eros's than Itokawa's.

- The observed reduced critical spin rate can be partially, but not completely, explained if the ejected body had a larger size when disruption occurred, and subsequently continued to break up to the present size.

- Introducing cohesion of hundreds of Pa allows the shape models to reach a critical rotation period of 2.2 h for all objects. Even for the most stable case, of spherical-shape asteroids, with density of 2.7 gr cm$^{-3}$ and a large secondary, some cohesion is required. A cohesionless structure for km-sized rubble pile asteroids is possible only if they disintegrate at a lower rotation rate with a period of 2.4-2.6 h and their density is near the higher range considered here.

Rotationally-disrupted pairs offer unique insights into understanding the internal structure of asteroids. Obtaining the shape models of additional primary components and secondary members through lightcurve inversion or other means, can provide further constraints on these bodies' hidden properties.

**Acknowledgments:** We thank Josef Durech and Josef Hanus for providing us with their lightcurve inversion code. DP is grateful to the ministry of Science and Technology of the Israeli government for their Ramon fellowship for post-docs, and for the Koshland Foundation for their partial founding. OA wishes to acknowledge the Helen Kimmel Center for Planetary

Science and Minerva Center program for support. We thank the Wise Observatory stuff for their continuous help.

**Plots**

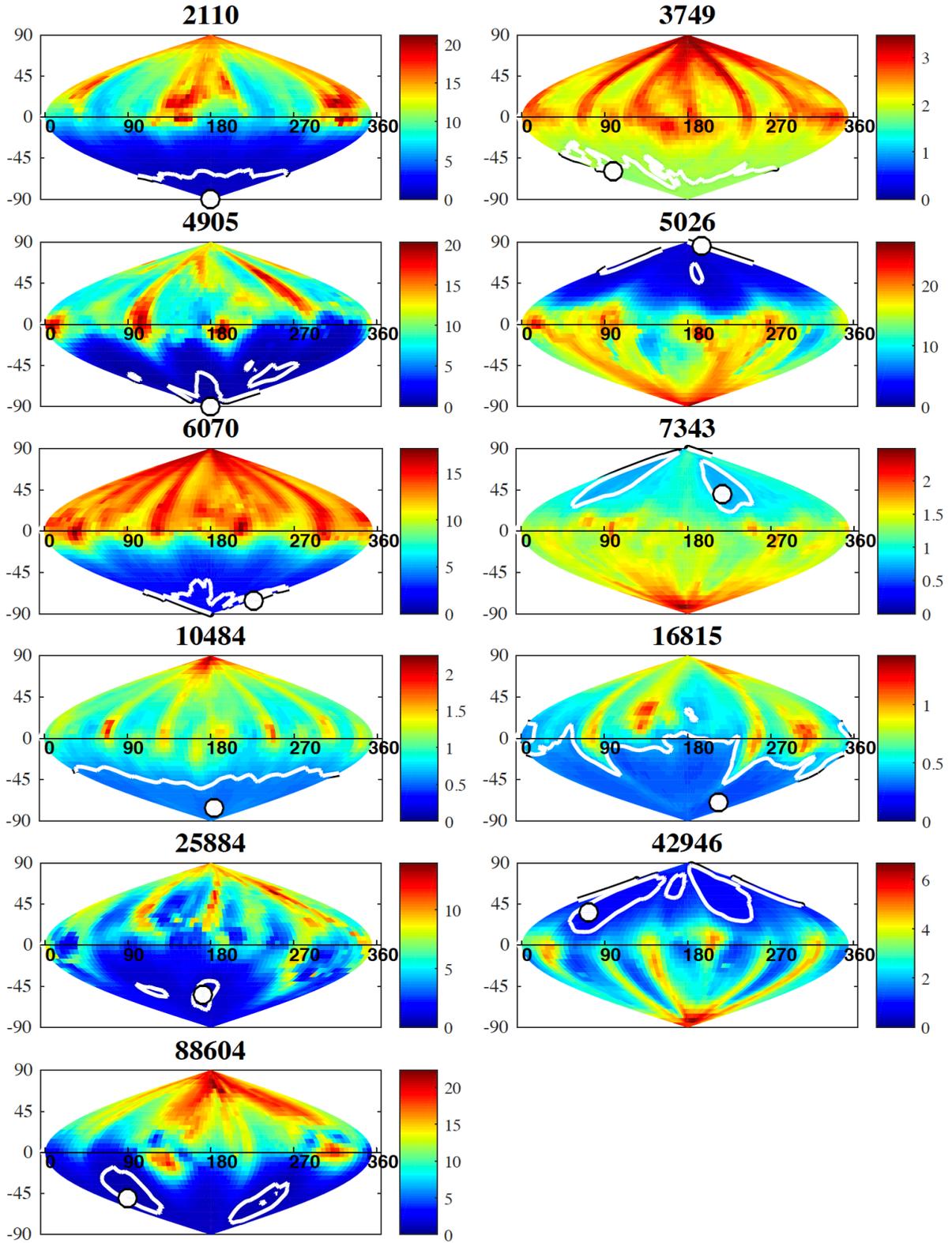

Fig. 1: The $\chi^2$ values for all spin axis solutions in ecliptic longitude–latitude coordinates for the eleven asteroids in our sample. The uncertainty of the fit corresponds to $3\sigma$ (white lines) above the global minimum, demonstrating the sense of rotation of the asteroids is prograde for 5026, 7343, 42946, and retrograde for 2110, 3749, 4905, 6070, 10484, 16815, 25884, 88604.

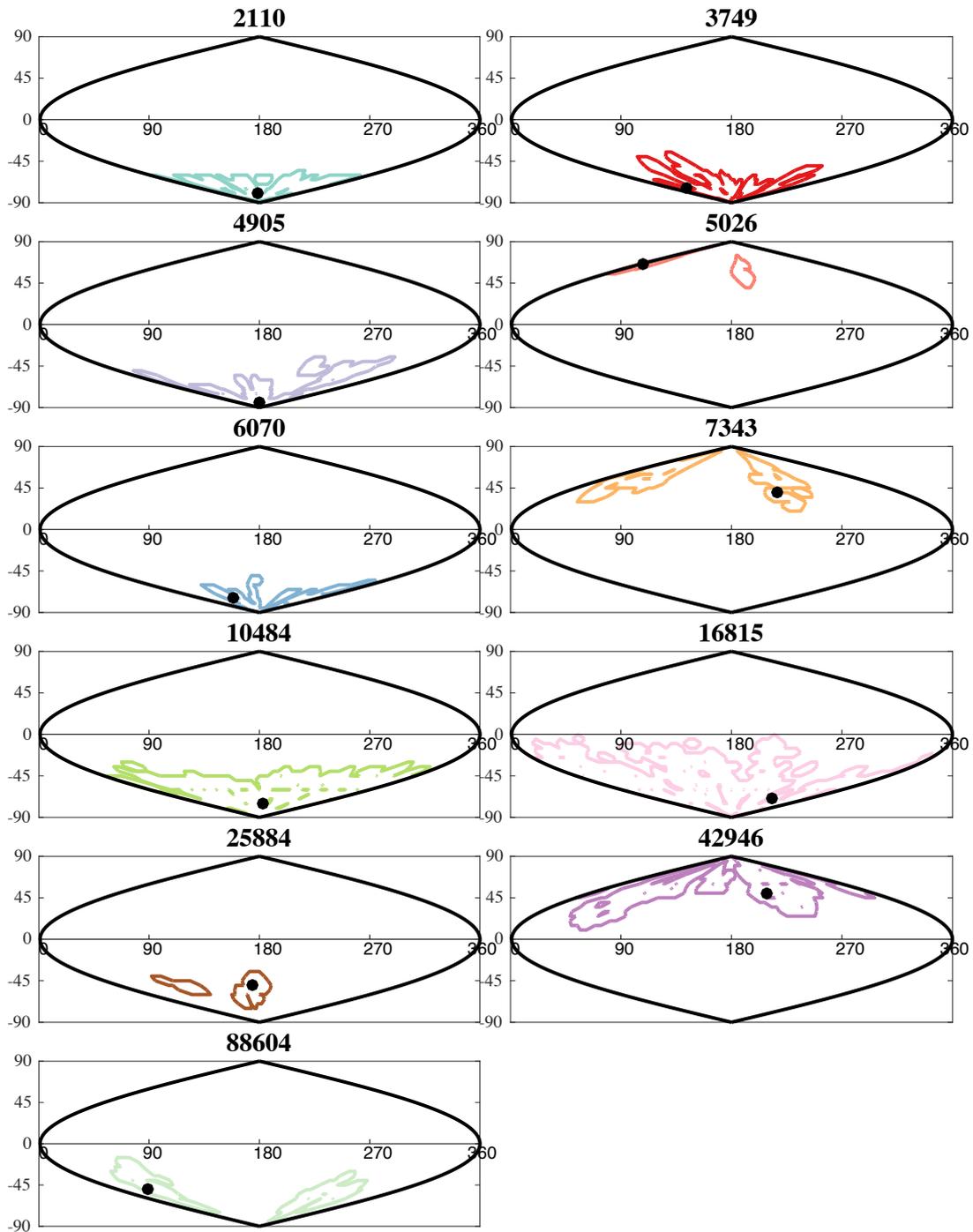

Fig. 2: Possible coordinates of spin axes in ecliptic longitude-latitude coordinates that have feasible physical parameters for each of the eleven asteroids in our sample. The model with lowest $\chi^2$ is marked with black circle for each object. This test further constrains the spin axis and shape models derived from the lightcurve inversion algorithm.

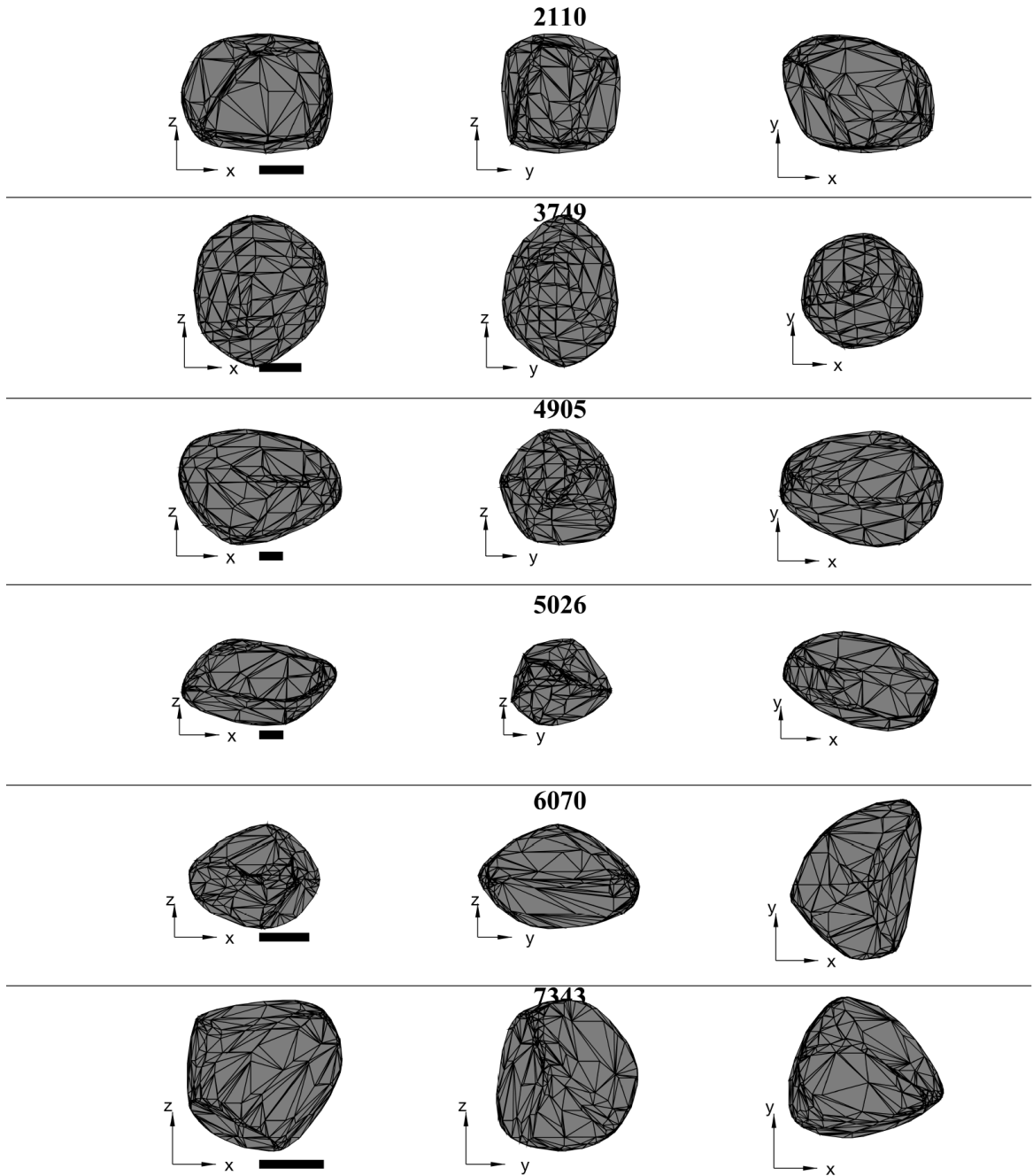

Fig. 3: Shape models for six asteroids in our sample. We present here the shape models that obtain the lowest $\chi^2$. Note that other shape models within the determined uncertainty for each asteroid are similar to the those presented here. The shape models are presented from two equatorial perspectives (left and center) and pole-on (right). The black scale bar is 1 km.

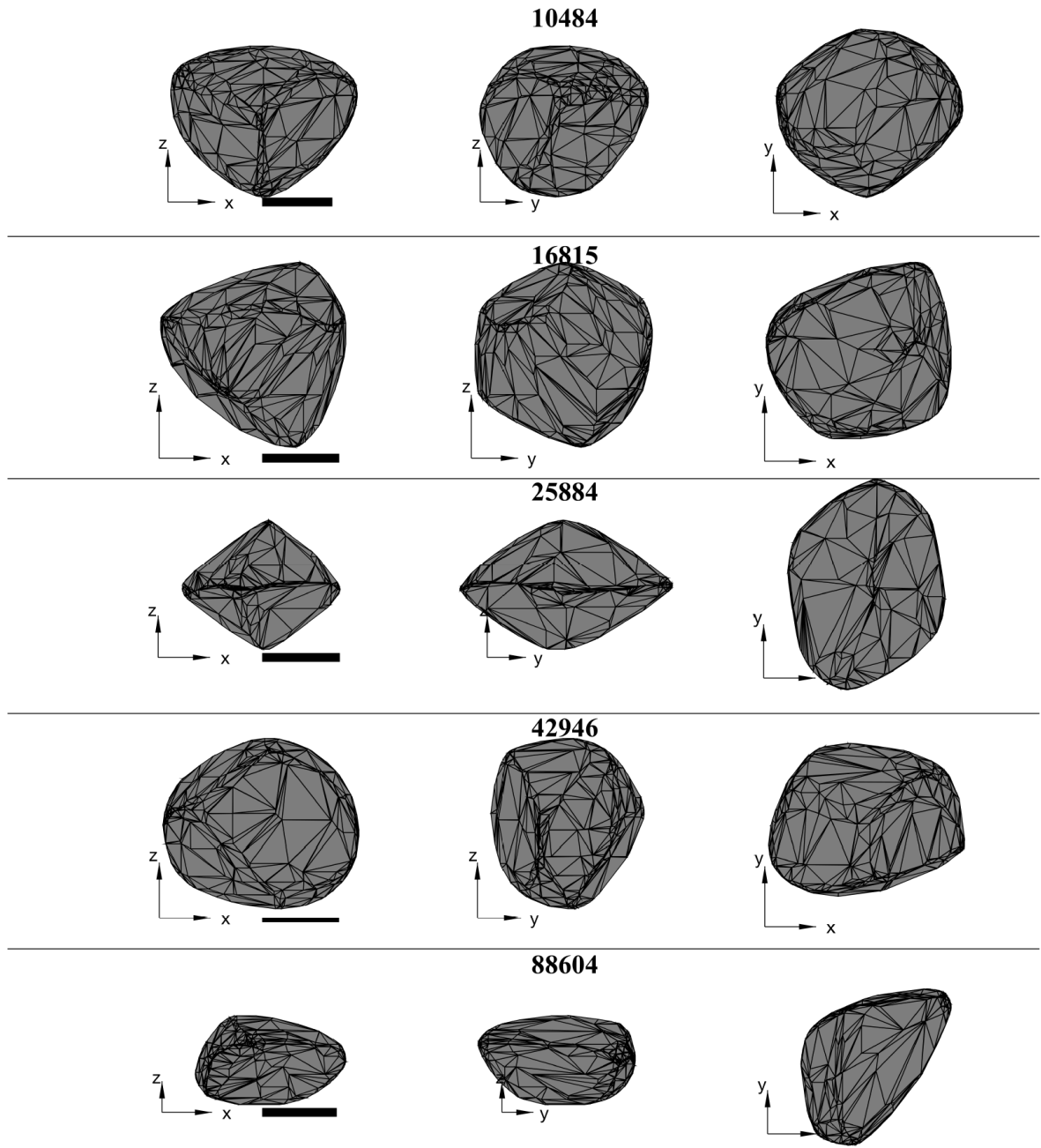

Fig. 4: Same as Fig. 3 for the additional five asteroids in our sample.

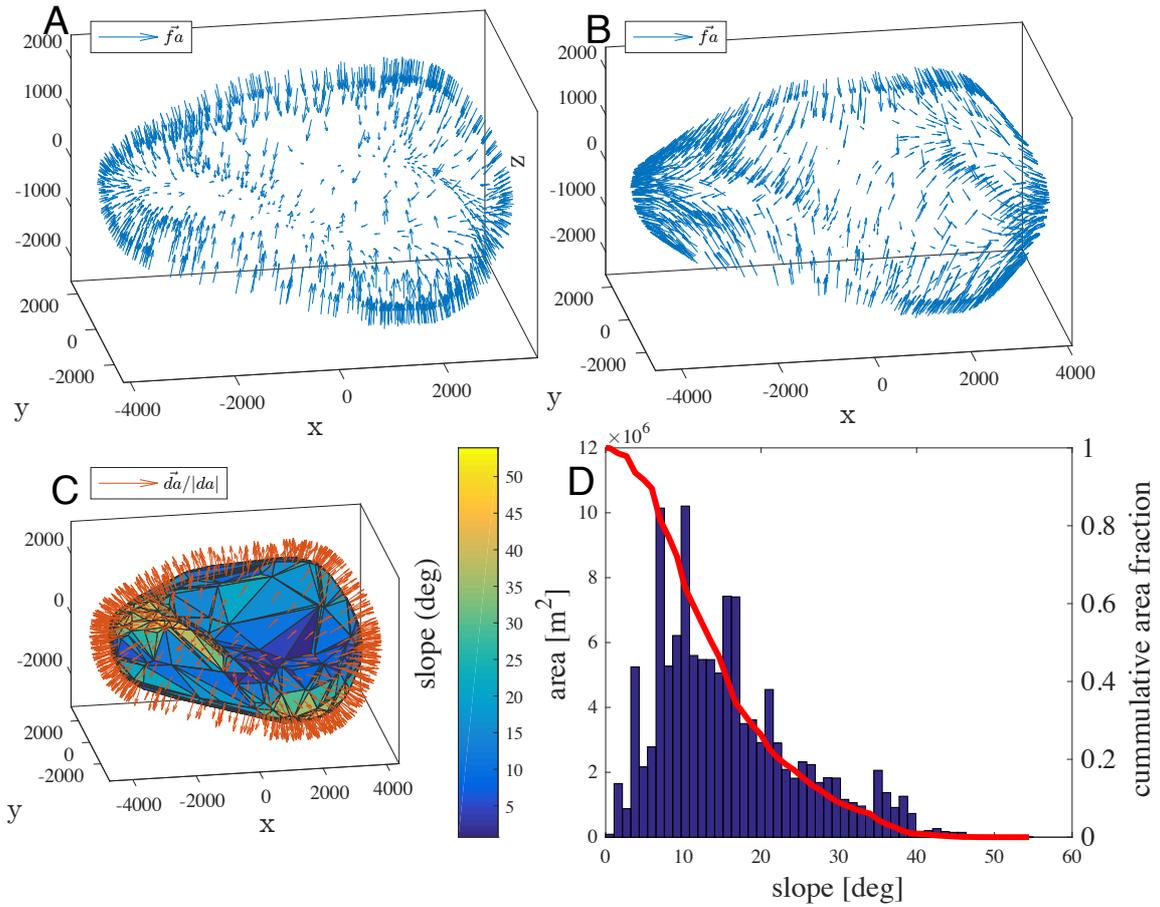

Fig. 5: An example of the surface slopes calculation. The sums of the gravitational and rotational acceleration vectors, for each facet, are drawn as blue arrows. In panel A, the shape model rotates slowly (3.34 h, the current rotation period of asteroid 2110 Moore-Sitterly), and the body's gravitation is stronger. In panel B, the same asteroid shape model is used but it rotates faster (2.2 h), and the rotational acceleration plays a greater role. Panel C presents the normal vectors to each facet (orange arrows), with the color scheme representing the computed slopes (for the slow rotating case). The distribution of the slopes is presented on panel D, while the red line is the cumulative area fraction.

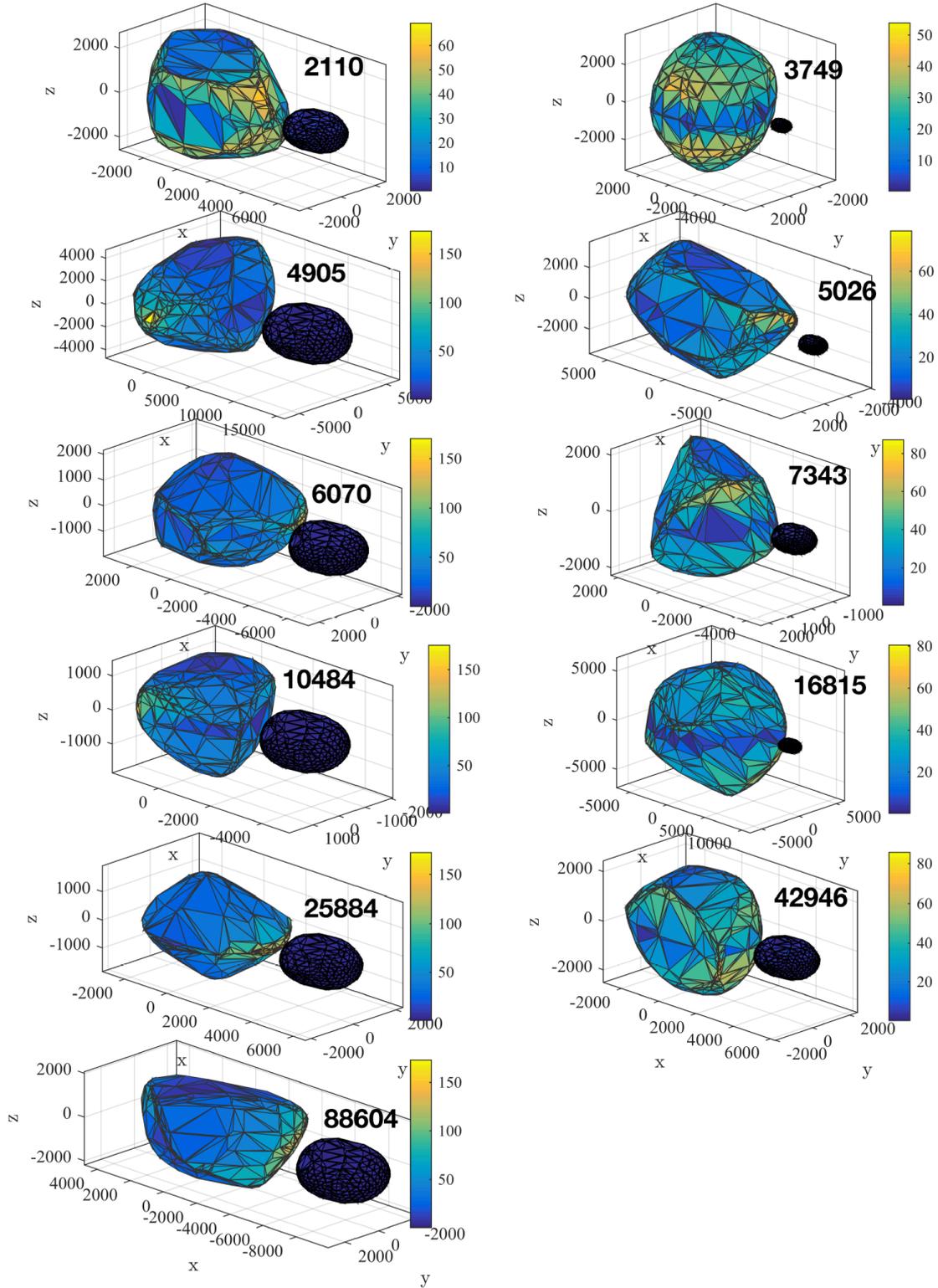

Fig. 6: Shape models of the eleven asteroids in our sample with slope values marked with the color bar. An additional shape at each panel represents the secondary, at the measured size ratio compared to the main asteroid. The secondary shape model shown is the same in all panels, and was derived using the lightcurve inversion technique using photometry measurements of asteroid 44612, the secondary of 2110 Moore-Sitterly (published in Polishook 2014a). The slope calculation presented here includes a density value of 2.0 gr cm-3 and the measured size ratio between each asteroid and its secondary.

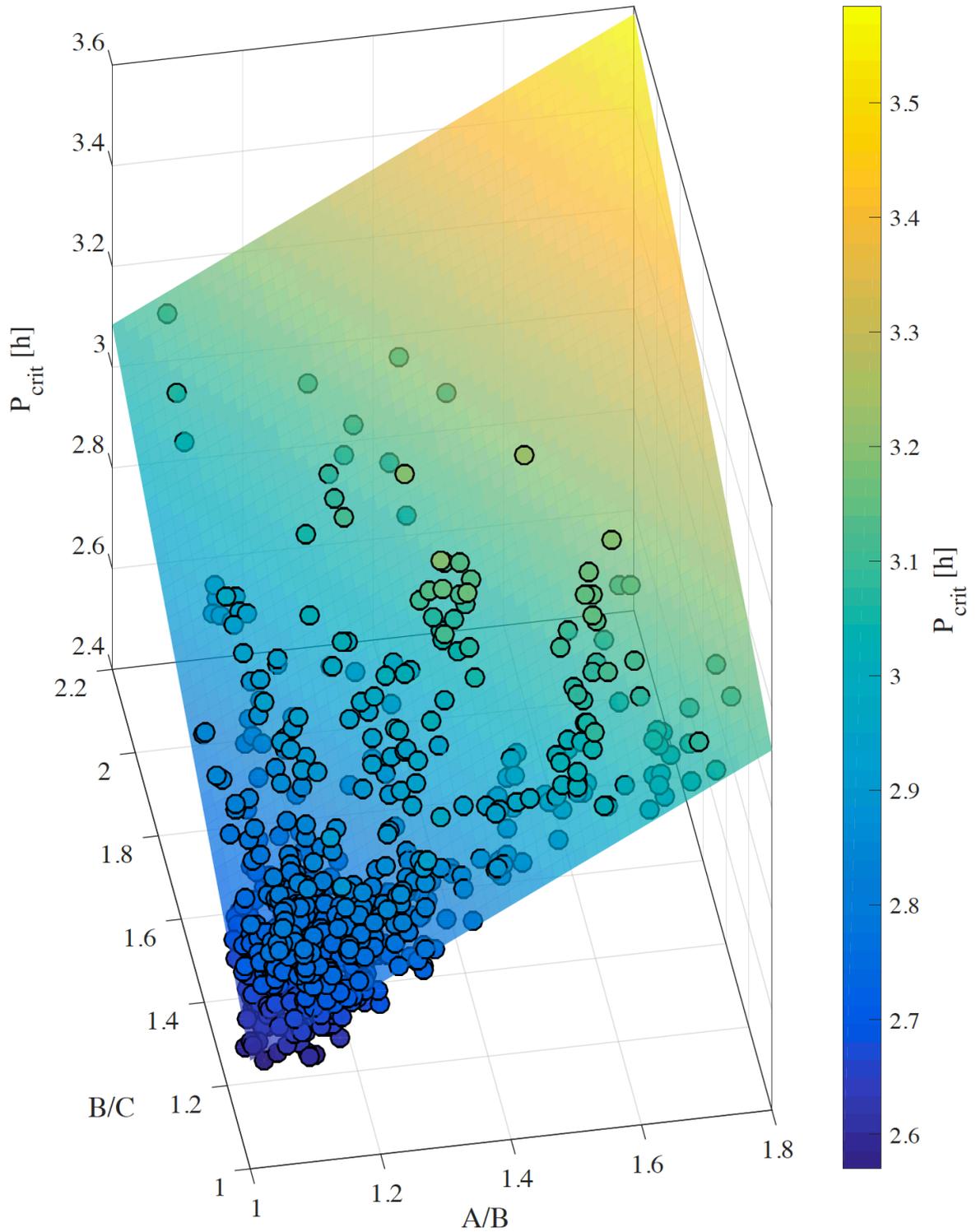

Fig. 7: The physical axes ratios ($A/B$ *vs.* $B/C$, when $A \geq B \geq C$), with the markers' colors representing the critical rotation period. Each circle represents a model of an asteroid pair from our sample. The partially transparent color scheme is a linear fit to both ratios, that shows that the critical rotation period is linearly correlated with the elongation and flattening. The plan has coefficients of $0.62 \pm 0.02$ ($A/B$ axis) and $0.39 \pm 0.02$ ($B/C$ axis).

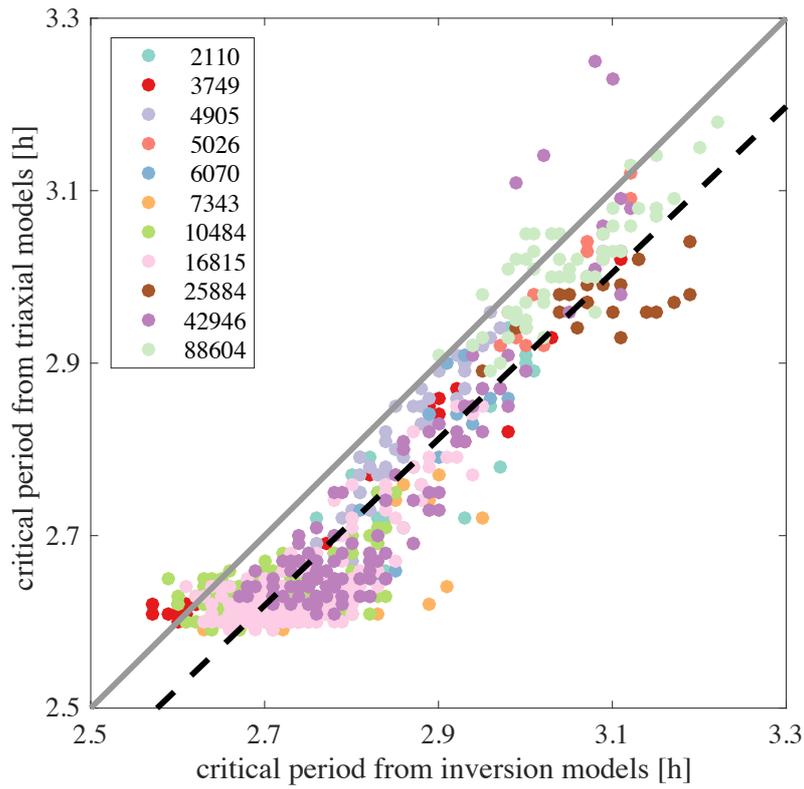

Fig. 8: Comparing the critical rotation periods for breakup using shape models that are derived from the inversion technique, to those that are derived from smooth triaxial shape models. Each color represents models of a single asteroid. While keeping the shape models constant, we normalized all other parameters to a single value (effective radius of $R$=3 km, a hypothetical secondary with a radius of 1.2 km, equivalent to magnitude difference $dH$=2, density $\rho$=2 g cm$^{-3}$; all are rounded values of the average of the real asteroids in our sample). The black dashed-line is a best linear fit to all data points with a slope of 0.96±0.14, while the grey line has a slope of 1 for comparison. Most (~95%) of the triaxial shape models disrupt at a somewhat faster rotation periods compared to the inversion-based shapes, with a median of 0.08 h and the 5th and the 95th percentile ranges between 0 to 0.17 h.

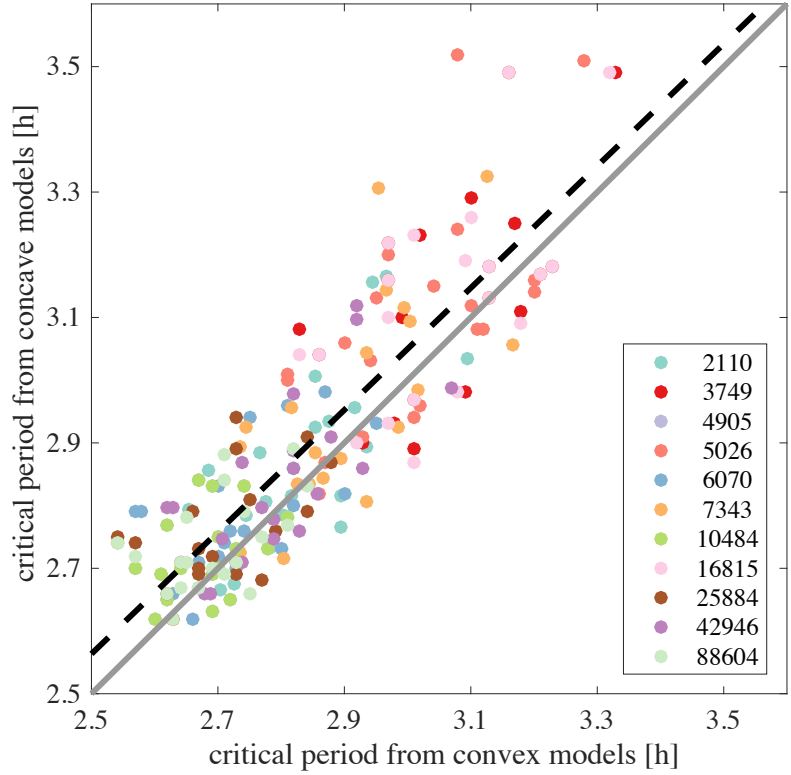

Fig. 9: Comparing the critical periods for breakup of convex and concave models. We perform our breakup calculation for a random selection of twenty asteroids from the DAMIT database, for which both convex and concave models were published. While keeping the shape parameters, we applied for each of the models the size parameters of all asteroids in our sample. The black dashed-line is a linear fit to the data points with a slope of 0.97±0.20, while the grey line has a slope of 1 for comparison. The convex models are slightly more stable with a median difference of 0.03 h towards the concave models and a 5th and a 95th percentile ranges between -0.085 to 0.215 h.

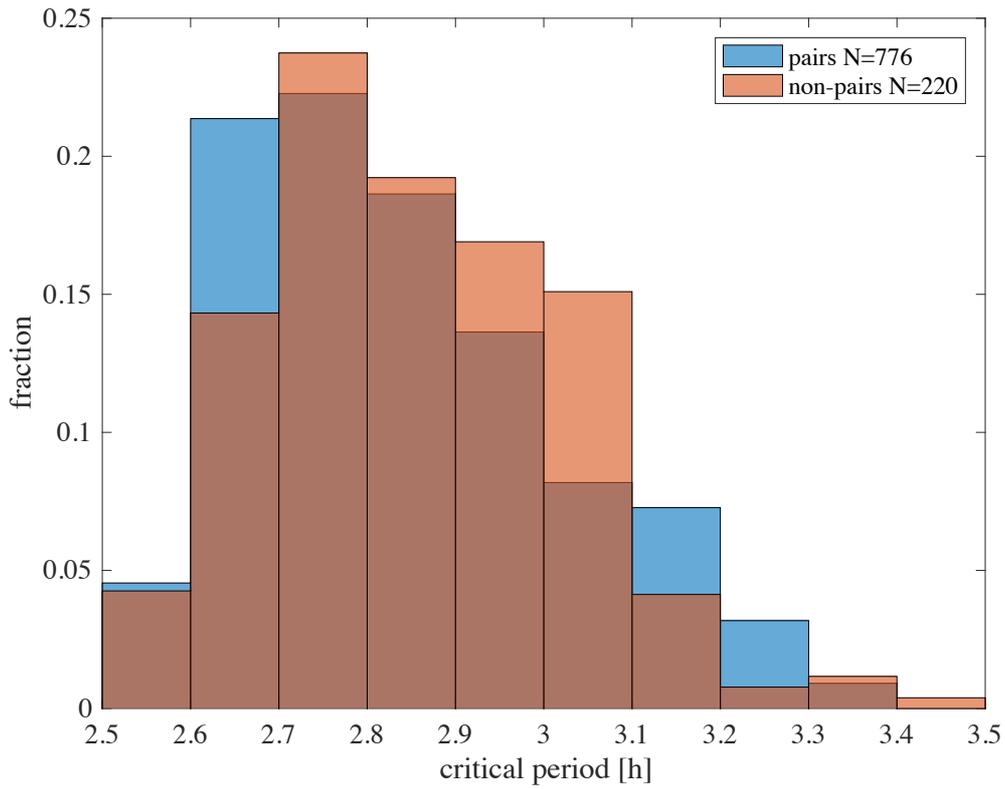

Fig. 10: Comparing the distributions of the critical rotation periods derived from the asteroid pairs' shape models (blue bars), and from convex shape models of other asteroids from the literature (pink bars) that are not recognize as pairs. All parameters were normalized to a single value, with only the shape models differed from to object to object. The two distributions are similar, suggesting that asteroid pairs are representative of non-pairs in terms of their shapes.

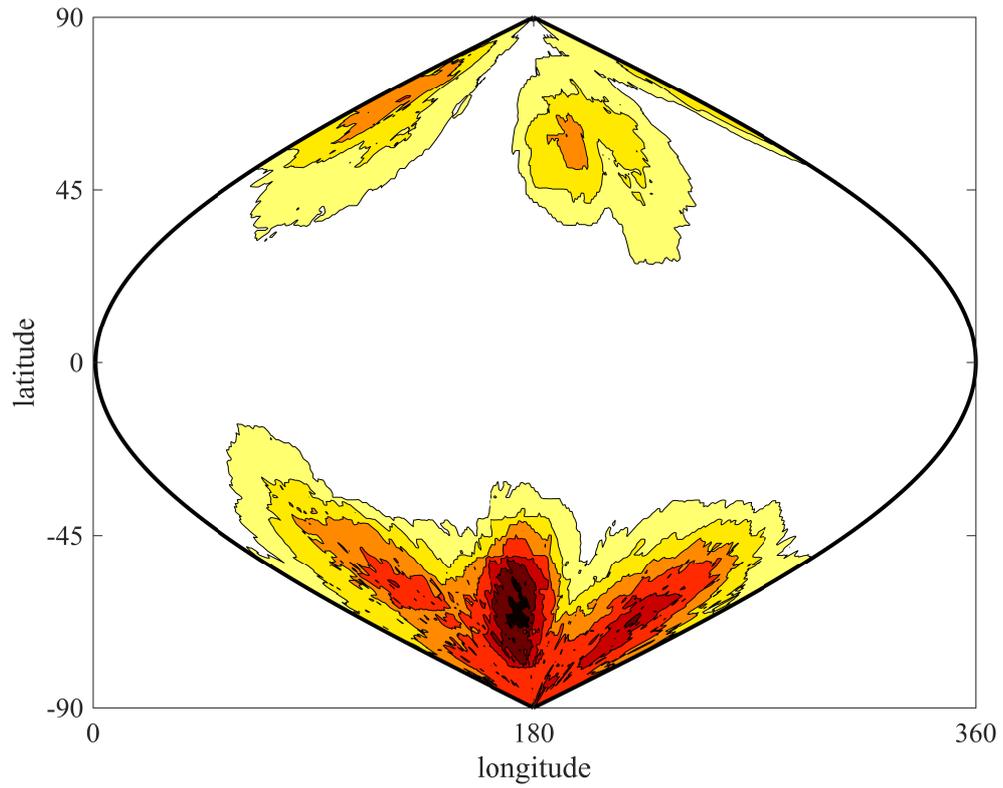

Fig. 11: A density map of the possible spin axis solutions averaged for all asteroids in our sample in a longitude-latitude plane. The concentration of the spin axes on the ecliptic poles is consistent with the YORP effect as a physical mechanism responsible to align the asteroids spin axis to their orbit.

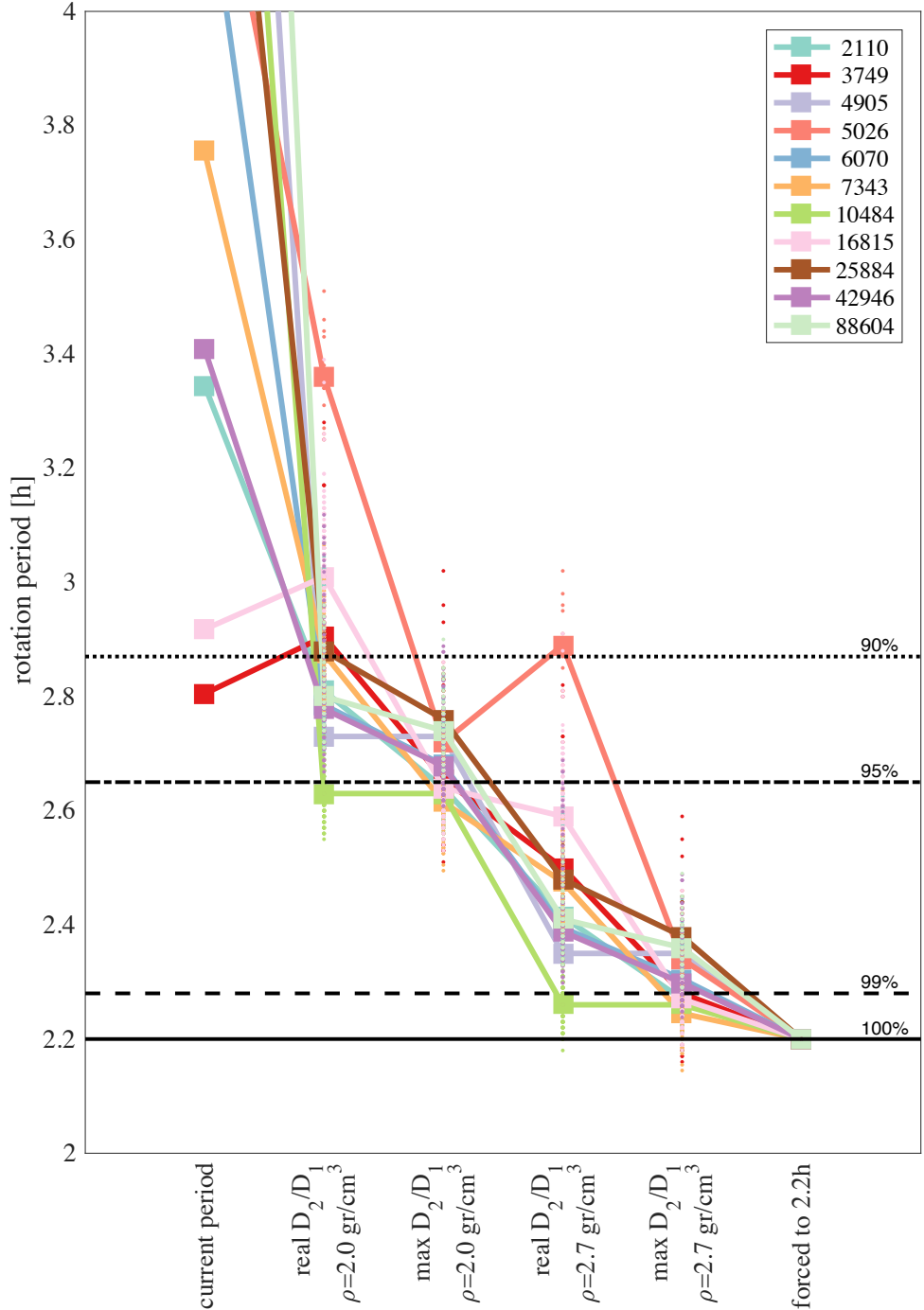

Fig. 12: Rotation periods of the eleven asteroid pairs in our sample determined in a suite of six scenarios from left to right: current rotation period (some values are off the plot in order to better see the models' results; see these periods at Table 2), critical rotation period for disruption with density of 2.0 gr cm⁻³ and current diameter ratio $D_2/D_1$, with density of 2.0 gr cm⁻³ and maximal diameter ratio $D_2/D_1 = 0.6$, with density of 2.7 gr cm⁻³ and current diameter ratio $D_2/D_1$, with density of 2.7 gr cm⁻³ and maximal diameter ratio $D_2/D_1 = 0.6$, and rotation period forced to the spin barrier at 2.2 h. The colored dots represent each shape model, while the colored squares represent the median of all shape models per asteroid. We compare our sample to all asteroids with known rotation period in the size range $1 < D < 40$ km, similar to the sizes of the studied pairs. The horizontal lines indicate the rotation period for which 90%, 95%, 99%, or 100% of the objects rotate slower than this value.

**Appendix**

Table A1: Observation circumstances. Additional observations that do not reported here, are available at Polishook 2014.

| Asteroid | Date | Telescope | Filter | Time span [h] | Num of Images | R [AU] | Delta [AU] | Phase angle [deg] | PAB long [deg] | PAB lat [deg] |
|---|---|---|---|---|---|---|---|---|---|---|
| 2110 | 20140527 | 0.46m | L | 2.16 | 31 | 1.88 | 1.38 | 31.87 | 308.1 | 0.7 |
| | 20140529 | 0.46m | L | 2.11 | 24 | 1.87 | 1.35 | 31.76 | 308.9 | 0.6 |
| | 20140603 | 0.46m | L | 1.44 | 25 | 1.87 | 1.3 | 31.39 | 310.8 | 0.6 |
| | 20140804 | 0.46m | L | 2.99 | 73 | 1.81 | 0.83 | 11.45 | 327.2 | 0 |
| | 20140805 | 0.46m | L | 6.18 | 147 | 1.81 | 0.82 | 10.8 | 327.4 | -0.1 |
| | 20140819 | 0.46m | L | 4.61 | 110 | 1.81 | 0.8 | 1.94 | 329 | -0.2 |
| | 20140820 | 0.46m | L | 3.08 | 62 | 1.81 | 0.8 | 1.28 | 329.2 | -0.3 |
| | 20140821 | 0.46m | L | 2.59 | 60 | 1.81 | 0.8 | 0.68 | 329.3 | -0.3 |
| | 20141120 | 0.46m | L | 4.37 | 67 | 1.86 | 1.44 | 31.72 | 352.4 | -1 |
| | 20141125 | 0.46m | L | 3.72 | 56 | 1.86 | 1.49 | 31.79 | 354.5 | -1 |
| | 20160104 | 0.71m | L | 1.61 | 24 | 2.55 | 1.73 | 14.92 | 138 | -0.2 |
| | 20160106 | 0.71m | L | 1.93 | 24 | 2.55 | 1.71 | 14.21 | 138.1 | -0.2 |
| | 20160116 | 0.71m | L | 6.61 | 59 | 2.56 | 1.64 | 10.22 | 138.3 | -0.2 |
| | 20160117 | 0.71m | L | 1.76 | 26 | 2.56 | 1.64 | 9.85 | 138.3 | -0.2 |
| | 20160204 | 0.71m | L | 2.61 | 34 | 2.57 | 1.58 | 1.38 | 138 | -0.1 |
| | 20160303 | 0.71m | L | 3.04 | 41 | 2.58 | 1.68 | 11.65 | 137.6 | 0.1 |
| | 20160304 | 0.71m | L | 3.01 | 40 | 2.58 | 1.69 | 12.04 | 137.7 | 0.1 |
| | 20160401 | 0.71m | L | 5.28 | 68 | 2.59 | 1.97 | 20.06 | 139.6 | 0.2 |
| | 20160402 | 0.71m | L | 4.96 | 65 | 2.59 | 1.98 | 20.24 | 139.8 | 0.2 |
| | 20170701 | 0.71m | L | 2.63 | 36 | 1.98 | 1 | 11.16 | 263.4 | 1.3 |
| | 20170703 | 0.71m | L | 3.02 | 40 | 1.97 | 1 | 12.3 | 263.5 | 1.3 |
| 3749 | 20140929 | 0.46m | L | 2.18 | 31 | 2.14 | 1.38 | 21.79 | 47.6 | 6.9 |
| | 20140930 | 0.46m | L | 1.16 | 17 | 2.14 | 1.37 | 21.47 | 47.8 | 6.9 |
| | 20141118 | 0.46m | L | 1.75 | 37 | 2.08 | 1.11 | 5.05 | 51.9 | 6.9 |
| | 20141119 | 0.46m | L | 2.3 | 36 | 2.08 | 1.11 | 5.37 | 51.9 | 6.9 |
| | 20141125 | 0.46m | L | 1.5 | 24 | 2.08 | 1.11 | 7.94 | 52.2 | 6.7 |
| | 20141228 | 0.46m | L | 0.96 | 15 | 2.05 | 1.29 | 22.4 | 55.8 | 5.3 |
| | 20150113 | 0.46m | L | 1.59 | 22 | 2.03 | 1.44 | 26.45 | 59.4 | 4.6 |
| | 20150119 | 0.46m | L | 1.96 | 28 | 2.03 | 1.5 | 27.48 | 61 | 4.3 |
| | 20150217 | 0.46m | L | 3.26 | 12 | 2.01 | 1.8 | 29.4 | 70.4 | 3 |
| | 20160304 | 0.71m | L | 0.62 | 9 | 2.27 | 1.81 | 24.93 | 227.1 | -6.2 |
| | 20160529 | 0.71m | L | 1.69 | 15 | 2.37 | 1.39 | 8.37 | 234.1 | -6.2 |
| | 20160602 | 0.71m | L | 0.71 | 11 | 2.37 | 1.41 | 10.08 | 234.2 | -6.1 |
| | 20160606 | 0.71m | L | 0.71 | 9 | 2.38 | 1.44 | 11.75 | 234.3 | -5.9 |
| | 20160626 | 1m | C | 2.03 | 29 | 2.4 | 1.61 | 18.69 | 235.9 | -5.3 |
| | 20160627 | 1m | C | 2.01 | 31 | 2.4 | 1.62 | 18.96 | 236 | -5.2 |
| | 20170925 | 0.71m | L | 1.59 | 26 | 2.3 | 1.31 | 5.07 | 356.2 | 6.6 |
| | 20171012 | 0.71m | L | 4.16 | 72 | 2.28 | 1.36 | 12.64 | 356.6 | 6.7 |
| | 20171013 | 0.71m | L | 4.49 | 72 | 2.28 | 1.37 | 13.04 | 356.7 | 6.7 |
| | 20171109 | 0.71m | L | 5.82 | 75 | 2.25 | 1.57 | 22.21 | 359.7 | 6.4 |
| | 20171110 | 0.71m | L | 2.43 | 36 | 2.25 | 1.58 | 22.43 | 359.8 | 6.4 |
| | 20171111 | 0.71m | L | 4.79 | 63 | 2.24 | 1.59 | 22.67 | 360 | 6.4 |
| 4905 | 20130927 | 0.46m | C | 3.11 | 61 | 2.17 | 1.21 | 10.98 | 23.5 | 0.1 |
| | 20131028 | 0.46m | C | 8.47 | 253 | 2.18 | 1.2 | 6.24 | 25.1 | -3 |
| | 20150224 | 0.46m | L | 7.16 | 115 | 2.89 | 1.92 | 4.82 | 152.1 | -11.2 |
| | 20150317 | 0.71m | L | 2.86 | 62 | 2.91 | 2.02 | 10.39 | 151.9 | -10.2 |
| | 20150324 | 0.46m | L | 5.73 | 74 | 2.92 | 2.08 | 12.46 | 152 | -9.8 |
| | 20150416 | 0.46m | L | 2.44 | 30 | 2.94 | 2.33 | 17.52 | 153.5 | -8.5 |
| | 20150421 | 0.46m | L | 3.57 | 36 | 2.95 | 2.4 | 18.23 | 154 | -8.2 |
| | 20150520 | 0.46m | L | 1.14 | 14 | 2.97 | 2.81 | 19.89 | 158.3 | -6.6 |
| | 20160402 | 0.71m | L | 3.79 | 34 | 2.97 | 2.15 | 13.01 | 228.4 | 6 |
| | 20160414 | 0.71m | L | 3.71 | 48 | 2.96 | 2.04 | 9.5 | 228.6 | 6.7 |
| | 20160415 | 0.71m | L | 2.94 | 39 | 2.96 | 2.04 | 9.19 | 228.7 | 6.7 |
| | 20160416 | 0.71m | L | 2.89 | 43 | 2.96 | 2.03 | 8.87 | 228.7 | 6.8 |
| | 20160429 | 0.71m | L | 2.49 | 40 | 2.95 | 1.96 | 4.7 | 228.5 | 7.5 |
| | 20160430 | 0.71m | L | 2.44 | 34 | 2.95 | 1.95 | 4.43 | 228.5 | 7.5 |
| | 20160703 | 1m | R | 2.35 | 43 | 2.87 | 2.25 | 18.25 | 229.4 | 9.3 |
| | 20160704 | 1m | R | 1.32 | 22 | 2.87 | 2.27 | 18.42 | 229.5 | 9.3 |
| | 20160711 | 0.71m | L | 2.20 | 33 | 2.86 | 2.34 | 19.42 | 230.3 | 9.4 |
| | 20160725 | 0.71m | L | 1.99 | 30 | 2.84 | 2.51 | 20.67 | 232.4 | 9.5 |
| | 20160727 | 0.71m | L | 1.99 | 28 | 2.84 | 2.53 | 20.77 | 232.7 | 9.5 |
| | 20170616 | 0.71m | L | 2.69 | 40 | 2.29 | 1.94 | 26.1 | 335.5 | 11.6 |
| | 20170617 | 0.71m | L | 2.7 | 38 | 2.29 | 1.93 | 26.08 | 335.8 | 11.6 |
| | 20170618 | 0.71m | L | 3.03 | 39 | 2.29 | 1.91 | 26.06 | 336 | 11.6 |
| | 20170619 | 0.71m | L | 3.02 | 44 | 2.29 | 1.9 | 26.03 | 336.3 | 11.6 |
| | 20170620 | 0.71m | L | 2.97 | 44 | 2.29 | 1.89 | 26 | 336.6 | 11.6 |
| | 20170629 | 0.71m | L | 3.41 | 50 | 2.27 | 1.78 | 25.49 | 339 | 11.6 |
| | 20170630 | 0.71m | L | 3.72 | 55 | 2.27 | 1.77 | 25.41 | 339.3 | 11.6 |
| | 20170701 | 0.71m | L | 2.68 | 39 | 2.27 | 1.75 | 25.32 | 339.6 | 11.5 |
| | 20170719 | 0.71m | L | 1.04 | 21 | 2.25 | 1.55 | 22.79 | 343.6 | 11.3 |
| | 20180111 | 0.71m | L | 0.69 | 11 | 2.17 | 2.25 | 25.59 | 16.3 | -2.4 |
| | 20181130 | 0.72m | L | 0.69 | 10 | 2.65 | 2.2 | 21 | 131.2 | -13.6 |
| | 20181215 | 0.72m | L | 2.88 | 42 | 2.67 | 2.05 | 18.68 | 133.3 | -14 |
| | 20181216 | 0.72m | L | 4.94 | 26 | 2.67 | 2.04 | 18.47 | 133.4 | -14.1 |
| | 20190115 | 0.72m | L | 2.68 | 32 | 2.73 | 1.83 | 10.24 | 135.0 | -14.5 |
| 5026 | 20150217 | 0.46m | L | 4.76 | 61 | 2.95 | 1.97 | 1.66 | 145 | -1.9 |
| | 20150225 | 0.46m | L | 4.91 | 54 | 2.95 | 1.98 | 4.87 | 144.7 | -2 |
| | 20160401 | 0.71m | L | 2.97 | 20 | 2.26 | 1.59 | 22.66 | 241.5 | -5.1 |
| | 20160529 | 0.71m | L | 0.62 | 8 | 2.1 | 1.09 | 3.35 | 248 | -5.2 |
| | 20160530 | 0.71m | L | 1.12 | 19 | 2.1 | 1.09 | 3.47 | 248 | -5.2 |
| | 20160531 | 0.71m | L | 1.18 | 19 | 2.1 | 1.09 | 3.67 | 248.1 | -5.2 |
| | 20160601 | 0.71m | L | 1.06 | 17 | 2.09 | 1.08 | 3.95 | 248.1 | -5.2 |
| | 20160602 | 0.71m | L | 1.00 | 17 | 2.09 | 1.08 | 4.29 | 248.1 | -5.2 |
| | 20160606 | 0.71m | L | 1.31 | 22 | 2.08 | 1.08 | 6.03 | 248.3 | -5.1 |
| | 20160607 | 0.71m | L | 1.13 | 19 | 2.08 | 1.08 | 6.52 | 248.4 | -5.1 |
| | 20171226 | 0.71m | L | 0.43 | 7 | 2.65 | 1.67 | 2.15 | 91.4 | 3 |
| | 20180107 | 0.71m | L | 5.38 | 34 | 2.67 | 1.73 | 7.21 | 91.1 | 2.7 |
| | 20180109 | 0.71m | L | 1.17 | 15 | 2.68 | 1.74 | 8.04 | 91.1 | 2.7 |
| 6070 | 20150319 | 0.71m | L | 2.41 | 46 | 2.88 | 1.89 | 3.79 | 187 | 3.9 |
| | 20150320 | 0.71m | L | 2.21 | 37 | 2.88 | 1.89 | 3.43 | 187 | 3.9 |
| | 20150324 | 0.71m | L | 2.00 | 36 | 2.88 | 1.88 | 2.07 | 186.8 | 3.9 |
| | 20150413 | 0.71m | L | 3.88 | 51 | 2.89 | 1.93 | 7.2 | 186.2 | 3.8 |
| | 20160626 | 1m | C | 4.47 | 71 | 2.34 | 1.33 | 3.17 | 281 | -1.2 |
| | 20160627 | 1m | C | 3.26 | 27 | 2.34 | 1.33 | 2.68 | 281.1 | -1.2 |
| | 20160703 | 1m | R | 3.64 | 54 | 2.33 | 1.31 | 0.91 | 281.1 | -1.3 |
| | 20160704 | 1m | R | 2.41 | 39 | 2.32 | 1.31 | 1.31 | 281.1 | -1.3 |
| | 20160710 | 0.71m | L | 2.87 | 26 | 2.31 | 1.3 | 4.33 | 281.2 | -1.4 |
| | 20160711 | 0.71m | L | 2.66 | 28 | 2.31 | 1.3 | 4.85 | 281.2 | -1.5 |
| | 20180107 | 0.71m | L | 6.79 | 50 | 2.38 | 1.42 | 7.29 | 121.4 | 2.3 |
| | 20180223 | 0.71m | L | 2.34 | 27 | 2.49 | 1.65 | 14.88 | 122.1 | 2.8 |
| 7343 | 20150520 | 0.46m | L | 0.62 | 5 | 2.49 | 1.48 | 2.27 | 238.1 | -4.3 |
| | 20150521 | 0.46m | L | 2.3 | 29 | 2.49 | 1.48 | 2.45 | 238.1 | -4.3 |
| | 20150522 | 0.46m | L | 0.6 | 8 | 2.49 | 1.48 | 2.7 | 238.1 | -4.3 |
| | 20150621 | 0.46m | L | 2.66 | 23 | 2.48 | 1.61 | 15.03 | 238.4 | -4.5 |
| | 20150622 | 0.46m | L | 1.62 | 7 | 2.48 | 1.62 | 15.39 | 238.5 | -4.5 |
| | 20150624 | 0.46m | L | 2.30 | 7 | 2.48 | 1.63 | 16.07 | 238.6 | -4.5 |
| | 20160910 | 0.71m | L | 5.77 | 84 | 1.92 | 1.18 | 26.55 | 33.4 | 1.4 |
| | 20160911 | 0.71m | L | 6.10 | 89 | 1.91 | 1.17 | 26.3 | 33.6 | 1.5 |
| | 20161103 | 1m | R | 4.59 | 109 | 1.89 | 0.9 | 2.65 | 41.7 | 3.7 |
| | 20161206 | 0.71m | L | 6.63 | 94 | 1.89 | 1.02 | 19.35 | 45.4 | 4.4 |
| 10484 | 20081125 | 0.46m | C | 5.90 | 44 | 2.14 | 1.43 | 22.54 | 108.9 | -3 |
| | 20081128 | 1m | R | 1.96 | 18 | 2.14 | 1.4 | 21.66 | 109.4 | -3.2 |
| | 20081129 | 1m | R | 5.64 | 20 | 2.14 | 1.39 | 21.39 | 109.6 | -3.2 |
| | 20081130 | 1m | R | 6.01 | 105 | 2.14 | 1.38 | 21.08 | 109.8 | -3.2 |
| | 20081201 | 1m | R | 5.33 | 95 | 2.14 | 1.37 | 20.75 | 109.9 | -3.3 |
| | 20111031 | 0.46m | C | 6.33 | 86 | 2.21 | 1.25 | 9.3 | 22.4 | 5.8 |
| | 20130405 | 0.46m | C | 2.62 | 31 | 2.35 | 1.38 | 6.58 | 185.7 | -7.1 |
| | 20130408 | 0.46m | C | 5.13 | 37 | 2.36 | 1.39 | 7.88 | 185.8 | -7 |
| | 20140724 | 0.46m | L | 4.63 | 44 | 2.45 | 1.44 | 4.19 | 306.9 | 6.1 |
| | 20140729 | 0.46m | L | 3.56 | 46 | 2.45 | 1.44 | 3.26 | 306.9 | 6.2 |
| | 20140731 | 0.46m | L | 6.18 | 67 | 2.44 | 1.44 | 3.37 | 306.9 | 6.3 |
| | 20140801 | 0.46m | L | 6.05 | 60 | 2.44 | 1.44 | 3.52 | 306.9 | 6.3 |
| | 20140804 | 0.46m | L | 1.61 | 21 | 2.44 | 1.44 | 4.28 | 306.9 | 6.3 |
| | 20170531 | 0.71m | L | 3.55 | 36 | 2.49 | 1.49 | 3.33 | 243.8 | -1 |
| | 20170616 | 0.71m | L | 3.00 | 34 | 2.5 | 1.55 | 10.61 | 244 | -0.5 |
| | 20170617 | 0.71m | L | 4.44 | 29 | 2.5 | 1.55 | 11.02 | 244 | -0.5 |
| | 20170620 | 0.71m | L | 4.29 | 38 | 2.5 | 1.57 | 12.23 | 244.1 | -0.4 |
| | 20180804 | 0.71m | L | 4.19 | 50 | 2.31 | 1.69 | 23.5 | 5.1 | 6.8 |
| | 20180816 | 0.71m | L | 5.93 | 70 | 2.3 | 1.55 | 21.06 | 7.3 | 7 |
| | 20180902 | 0.72m | L | 5.63 | 67 | 2.28 | 1.4 | 15.82 | 9.4 | 7.2 |
| | 20180903 | 0.72m | L | 0.61 | 9 | 2.28 | 1.39 | 15.48 | 9.5 | 7.2 |
| | 20181004 | 0.72m | L | 1.34 | 23 | 2.25 | 1.26 | 4.03 | 10.9 | 7.1 |
| | 20181102 | 0.72m | L | 6.39 | 85 | 2.23 | 1.35 | 15.19 | 12.2 | 6.2 |
| | 20181116 | 0.72m | L | 3.80 | 45 | 2.22 | 1.45 | 20.16 | 13.9 | 5.6 |
| | 20181117 | 0.72m | L | 5.53 | 57 | 2.22 | 1.46 | 20.46 | 14 | 5.5 |
| | 20181130 | 0.72m | L | 5.47 | 70 | 2.21 | 1.59 | 23.65 | 16.4 | 5 |
| | 20181201 | 0.72m | L | 5.49 | 68 | 2.21 | 1.6 | 23.84 | 16.6 | 4.9 |
| 16815 | 20131030 | 0.46m | C | 6.85 | 91 | 2.62 | 1.69 | 9.63 | 17.7 | -8.5 |
| | 20150113 | 0.46m | L | 1.41 | 20 | 2.53 | 1.61 | 9.72 | 129.5 | 13.2 |
| | 20150119 | 0.46m | L | 4.61 | 67 | 2.53 | 1.58 | 7.96 | 129.6 | 13.4 |
| | 20150122 | 0.46m | L | 1.19 | 19 | 2.53 | 1.58 | 7.29 | 129.6 | 13.5 |
| | 20150221 | 0.46m | L | 2.64 | 47 | 2.52 | 1.63 | 12.15 | 130 | 13.8 |
| | 20150315 | 0.46m | L | 1.56 | 11 | 2.52 | 1.8 | 18.63 | 131.8 | 13.3 |
| | 20150316 | 0.46m | L | 1.99 | 21 | 2.52 | 1.81 | 18.82 | 131.9 | 13.3 |
| | 20150319 | 0.46m | L | 1.68 | 16 | 2.52 | 1.84 | 19.54 | 132.3 | 13.2 |
| | 20150324 | 0.72m | L | 5.52 | 91 | 2.52 | 1.89 | 20.5 | 132.9 | 13.1 |
| | 20150413 | 0.46m | L | 3.94 | 50 | 2.51 | 2.12 | 23.05 | 136.6 | 12.3 |
| | 20150414 | 0.46m | L | 3.74 | 47 | 2.51 | 2.13 | 23.13 | 136.8 | 12.3 |
| | 20150512 | 0.46m | L | 2.44 | 24 | 2.51 | 2.47 | 23.41 | 143.8 | 11.3 |
| | 20150513 | 0.46m | L | 1.87 | 12 | 2.51 | 2.48 | 23.36 | 144 | 11.3 |
| | 20150514 | 0.46m | L | 2.94 | 26 | 2.51 | 2.49 | 23.32 | 144.3 | 11.2 |
| | 20180107 | 0.72m | L | 3.05 | 39 | 2.61 | 2.58 | 21.83 | 19 | -5.5 |
| | 20181031 | 0.72m | L | 3.75 | 50 | 2.55 | 2.25 | 22.73 | 110.5 | 7.9 |
| | 20181101 | 0.72m | L | 4.01 | 58 | 2.55 | 2.24 | 22.69 | 110.8 | 7.9 |
| | 20181102 | 0.72m | L | 4.10 | 60 | 2.55 | 2.22 | 22.64 | 111 | 8 |
| | 20181117 | 0.72m | L | 0.41 | 6 | 2.55 | 2.03 | 21.33 | 113.9 | 8.9 |
| | 20181130 | 0.72m | L | 2.96 | 44 | 2.55 | 1.88 | 19.13 | 115.9 | 9.8 |
| | 20190114 | 0.72m | L | 9.27 | 57 | 2.54 | 1.58 | 6.21 | 118.4 | 12.4 |
| 25884 | 20100320 | 1m | C | 1.54 | 26 | 2.07 | 1.14 | 13.44 | 174.1 | 21.4 |
| | 20100321 | 1m | C | 3.40 | 44 | 2.07 | 1.14 | 13.52 | 174.1 | 21.2 |
| | 20100409 | 1m | C | 7.20 | 108 | 2.05 | 1.18 | 18.19 | 174.6 | 18.2 |
| | 20100410 | 1m | C | 5.24 | 91 | 2.05 | 1.18 | 18.5 | 174.7 | 18.1 |
| | 20111028 | 0.46m | C | 4.83 | 69 | 1.94 | 0.95 | 6.29 | 44.1 | 1.6 |
| | 20130515 | 0.46m | C | 2.56 | 44 | 1.93 | 0.95 | 10.08 | 221.2 | -4.7 |
| | 20130519 | 0.46m | C | 2.27 | 27 | 1.93 | 0.96 | 12.62 | 221.4 | -5.5 |
| | 20130603 | 0.46m | C | 0.86 | 15 | 1.91 | 1.03 | 20.85 | 222.4 | -8.1 |
| | 20130605 | 0.46m | C | 2.87 | 31 | 1.91 | 1.05 | 21.74 | 222.7 | -8.5 |
| | 20130606 | 0.46m | C | 2.90 | 44 | 1.91 | 1.05 | 22.18 | 222.8 | -8.6 |
| | 20130607 | 0.46m | C | 2.83 | 46 | 1.91 | 1.06 | 22.62 | 222.9 | -8.8 |
| | 20141223 | 0.71m | L | 5.68 | 62 | 2.07 | 1.17 | 14.72 | 91.6 | 25 |
| | 20141224 | 0.71m | L | 8.49 | 101 | 2.07 | 1.17 | 14.77 | 91.6 | 25.1 |
| | 20141225 | 0.71m | L | 2.99 | 46 | 2.07 | 1.18 | 14.84 | 91.6 | 25.1 |
| | 20180206 | 0.71m | L | 8.93 | 129 | 2.11 | 1.24 | 16.7 | 139.5 | 29.6 |
| 42946 | 20130118 | 0.46m | C | 3.10 | 36 | 2.6 | 1.81 | 15.43 | 81.9 | -4 |
| | 20130119 | 0.46m | C | 4.37 | 63 | 2.6 | 1.82 | 15.7 | 82 | -4 |
| | 20140228 | 0.46m | C | 3.48 | 50 | 2.39 | 1.67 | 19.73 | 204.4 | 5.1 |
| | 20140303 | 0.46m | C | 3.78 | 44 | 2.39 | 1.64 | 18.97 | 204.9 | 5.2 |
| | 20140305 | 0.46m | C | 3.71 | 37 | 2.39 | 1.62 | 18.43 | 205.1 | 5.2 |
| | 20140307 | 0.46m | C | 3.80 | 40 | 2.39 | 1.6 | 17.86 | 205.4 | 5.3 |
| | 20140408 | 0.46m | C | 6.85 | 92 | 2.39 | 1.4 | 5.66 | 207.7 | 6 |
| | 20150811 | 0.46m | L | 5.83 | 60 | 2.68 | 1.69 | 5.34 | 330.2 | -1.5 |
| | 20150918 | 0.46m | L | 6.11 | 39 | 2.7 | 1.79 | 10.89 | 330.3 | -2.3 |
| | 20150920 | 0.46m | L | 3.11 | 34 | 2.7 | 1.8 | 11.59 | 330.4 | -2.3 |
| | 20161206 | 0.72m | L | 2.00 | 28 | 2.66 | 1.69 | 3.72 | 68.8 | -5.4 |
| | 20161207 | 0.72m | L | 1.58 | 24 | 2.66 | 1.69 | 4.08 | 68.8 | -5.4 |
| | 20161222 | 0.72m | L | 3.56 | 46 | 2.65 | 1.74 | 9.9 | 68.9 | -5.1 |
| | 20170104 | 0.72m | L | 6.82 | 62 | 2.65 | 1.84 | 14.53 | 69.5 | -4.8 |
| | 20170106 | 0.72m | L | 5.41 | 49 | 2.64 | 1.85 | 15.14 | 69.6 | -4.8 |
| | 20180208 | 0.72m | L | 5.75 | 67 | 2.39 | 1.72 | 20.46 | 187.5 | 4 |
| | 20180224 | 0.72m | L | 4.48 | 36 | 2.39 | 1.56 | 16.07 | 189.7 | 4.5 |
| | 20180704 | 0.72m | L | 1.66 | 20 | 2.4 | 2.17 | 25.1 | 205 | 5 |
| 88604 | 20090617 | 1m | R | 3.40 | 38 | 2.95 | 1.99 | 7.85 | 286.1 | -1.1 |
| | 20090618 | 1m | R | 3.24 | 35 | 2.95 | 1.98 | 7.49 | 286.1 | -1.1 |
| | 20090622 | 1m | R | 4.87 | 54 | 2.95 | 1.97 | 5.99 | 286.1 | -0.9 |
| | 20140904 | 0.46m | L | 2.86 | 32 | 2.85 | 1.86 | 5.26 | 339.5 | 12.1 |
| | 20140918 | 0.46m | L | 6.64 | 67 | 2.84 | 1.89 | 8.45 | 339.4 | 12.4 |
| | 20140929 | 0.46m | L | 4.34 | 50 | 2.83 | 1.95 | 11.79 | 339.5 | 12.5 |
| | 20141002 | 0.46m | L | 4.51 | 44 | 2.82 | 1.97 | 12.67 | 339.6 | 12.5 |
| | 20151119 | 0.46m | L | 2.33 | 33 | 2.42 | 1.61 | 16.5 | 92.4 | 5.7 |
| | 20151204 | 0.71m | L | 4.30 | 30 | 2.41 | 1.49 | 10.85 | 93.6 | 4.9 |
| | 20151205 | 0.71m | L | 8.85 | 63 | 2.41 | 1.48 | 10.39 | 93.6 | 4.8 |
| | 20151218 | 0.71m | L | 0.53 | 7 | 2.4 | 1.43 | 4.37 | 94 | 3.9 |
| | 20160104 | 0.71m | L | 3.02 | 25 | 2.4 | 1.43 | 4.93 | 94.3 | 2.7 |
| | 20160116 | 0.71m | L | 7.87 | 32 | 2.39 | 1.47 | 10.58 | 94.7 | 1.7 |
| | 20160117 | 0.71m | L | 7.66 | 51 | 2.39 | 1.48 | 11.01 | 94.8 | 1.7 |
| | 20160204 | 0.71m | L | 2.67 | 23 | 2.39 | 1.61 | 17.82 | 96.3 | 0.3 |
| | 20160301 | 0.71m | L | 3.94 | 29 | 2.38 | 1.89 | 23.33 | 100.7 | -1.4 |
| | 20180621 | 0.71m | L | 3.25 | 32 | 2.95 | 2.15 | 14.18 | 308.4 | 3.7 |
| | 20180622 | 0.71m | L | 4.08 | 60 | 2.95 | 2.14 | 13.93 | 308.4 | 3.8 |
| | 20180704 | 0.72m | L | 5.16 | 56 | 2.95 | 2.04 | 10.46 | 308.8 | 4.4 |

Lightcurves of the eleven asteroids in our sample. Photometry from each apparition is presented separately.

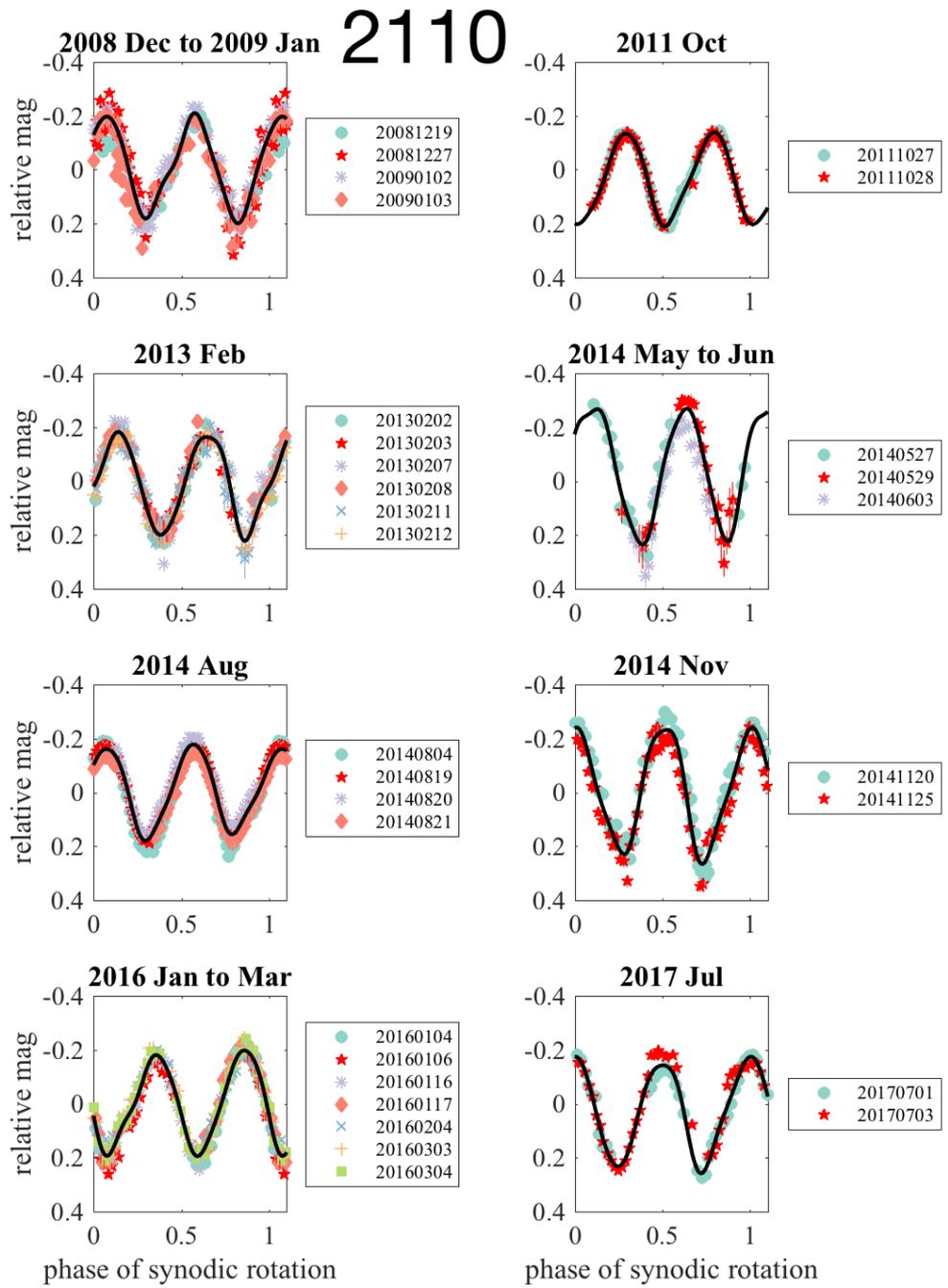

Fig. A1: Folded lightcurves of 2110 Moore-Sitterly using period of 3.34474 h.

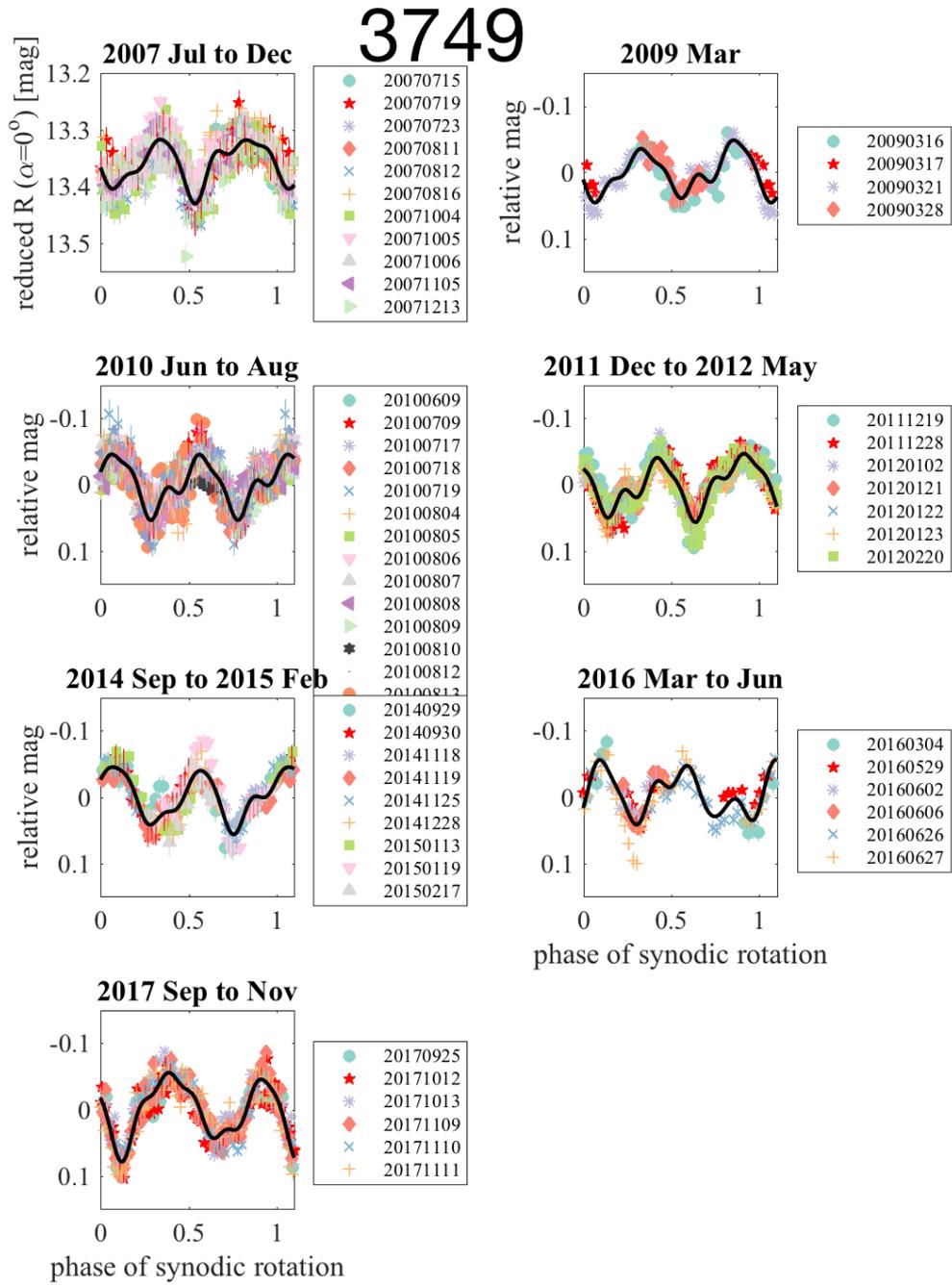

Fig. A2: Folded lightcurves of 3749 Balam using period of 2.80490 h.

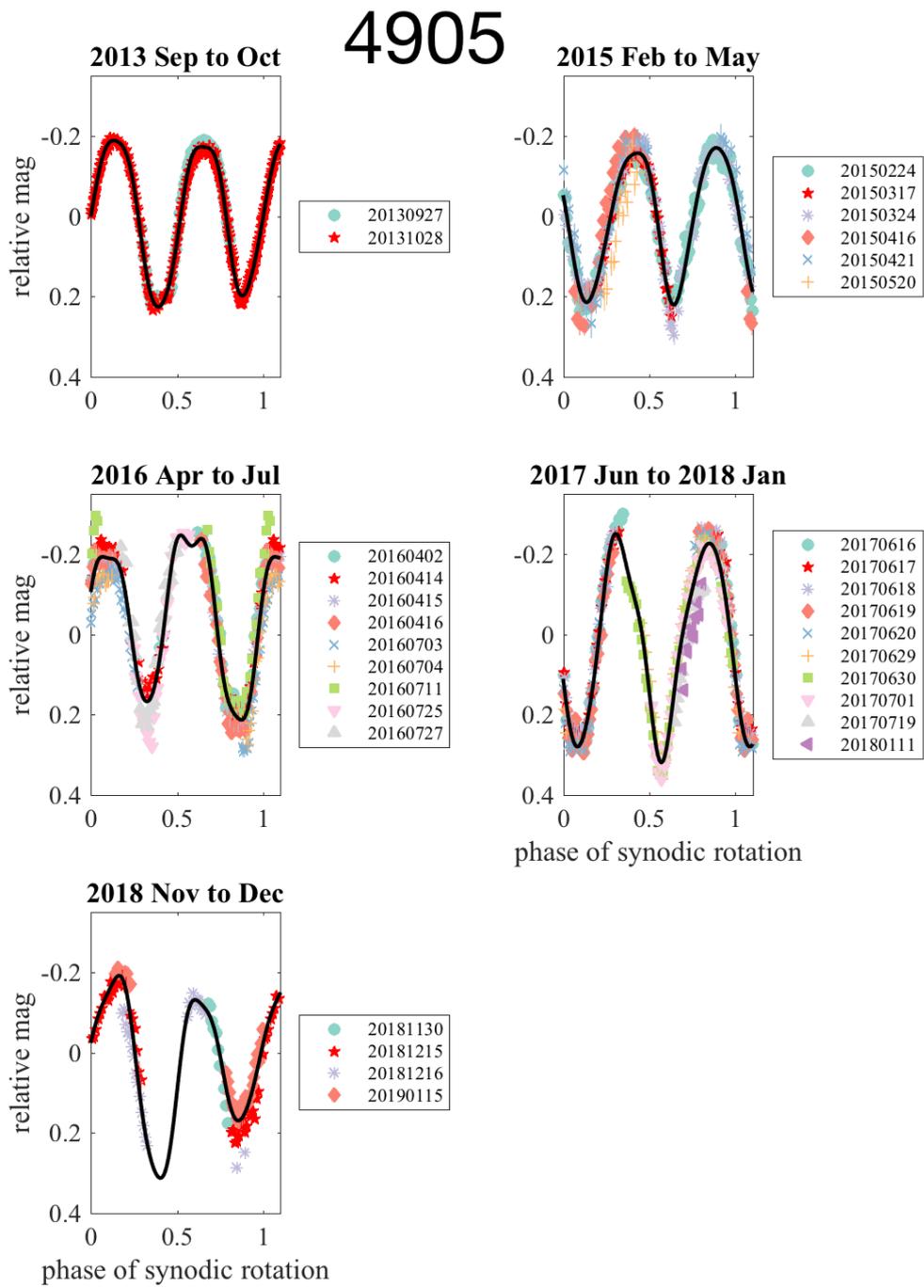

Fig. A3: Folded lightcurves of 4905 Hiromi using period of 6.0442 h.

# 5026

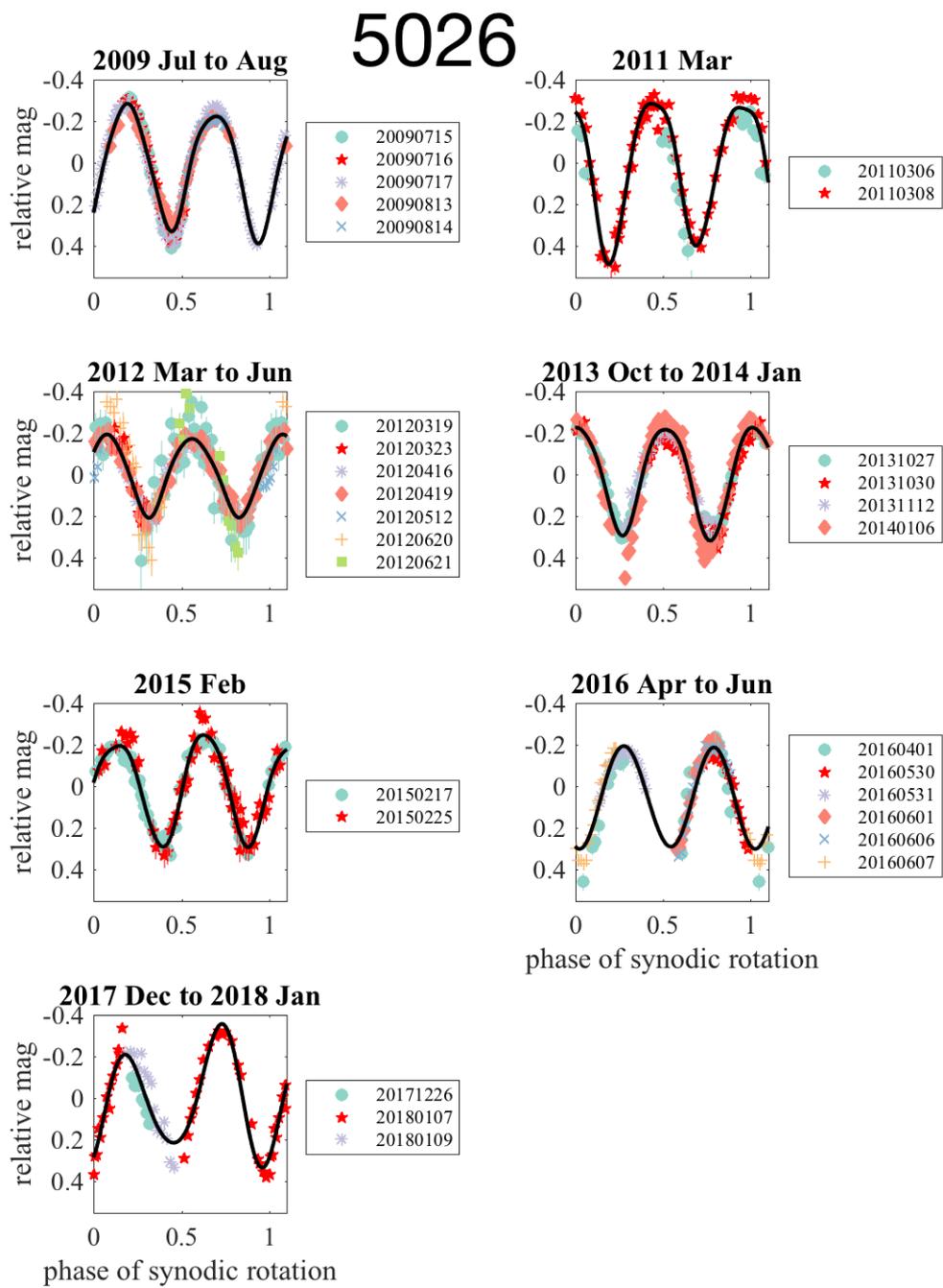

Fig. A4: Folded lightcurves of 5026 Martes using period of 4.4243 h.

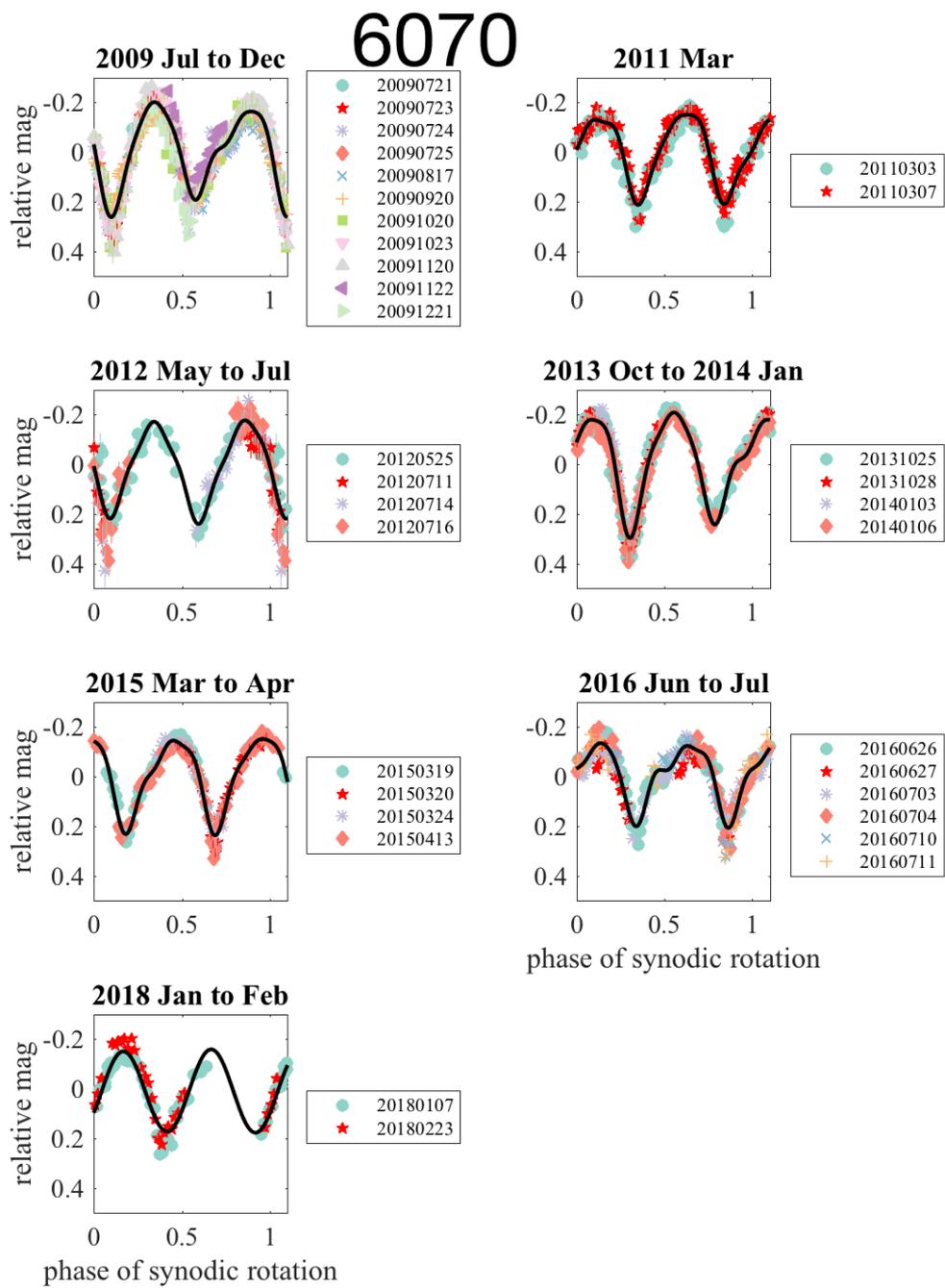

Fig. A5: Folded lightcurves of 6070 Rheinland using period of 4.2733 h.

# 7343

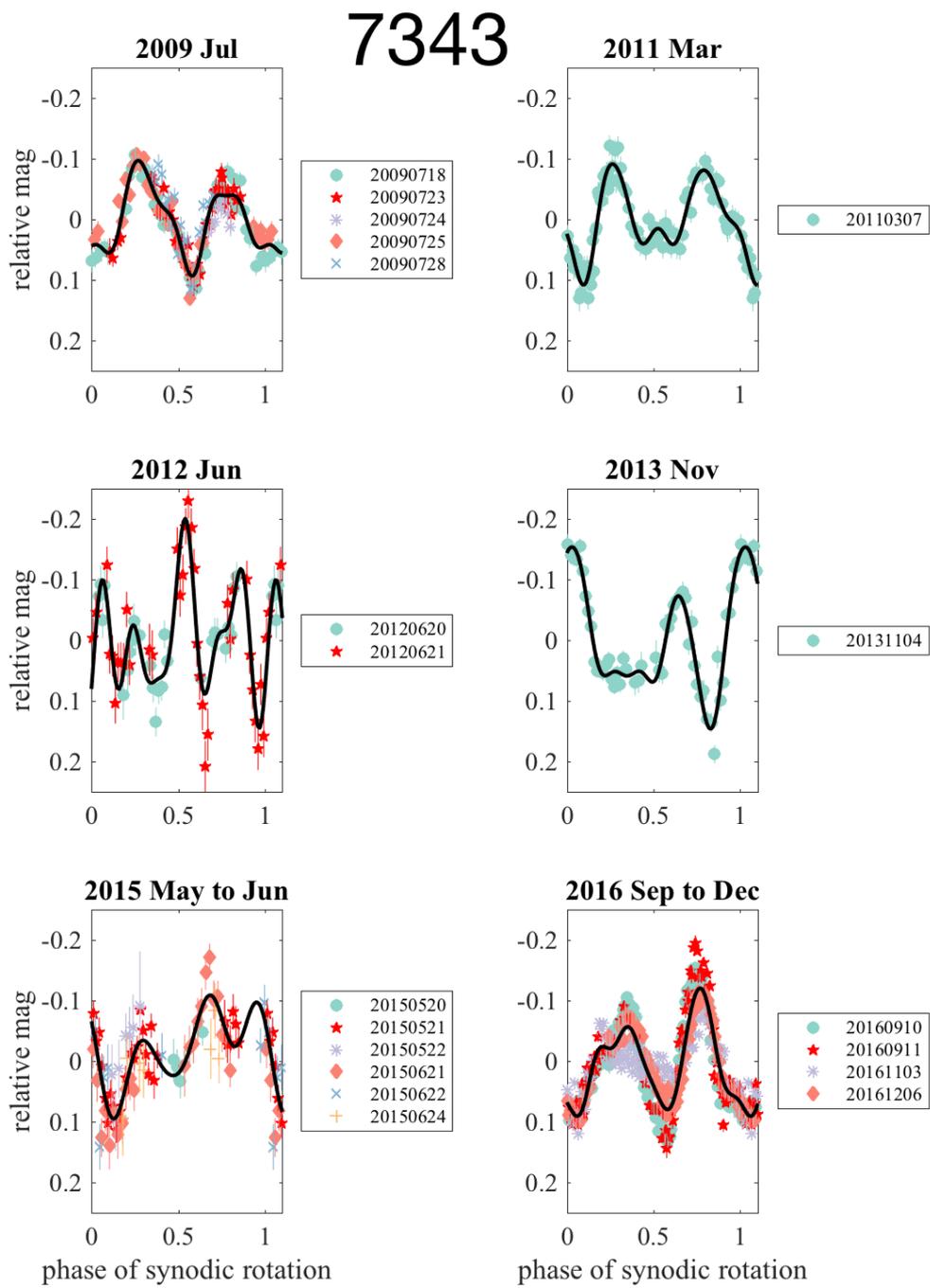

Fig. A6: Folded lightcurves of 7343 Ockeghem using period of 3.755 h.

# 10484

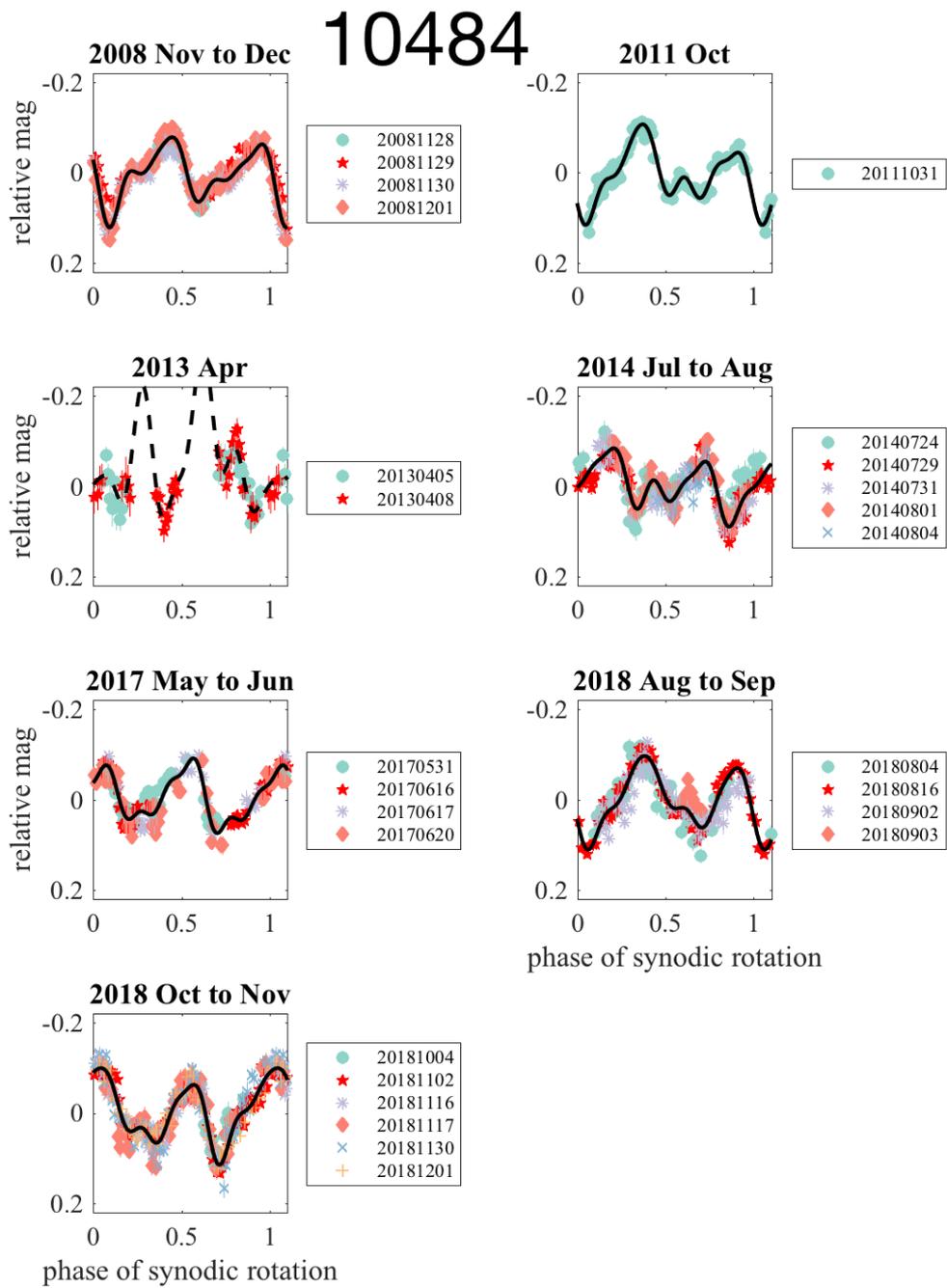

Fig. A7: Folded lightcurves of 10484 Hecht using period of 5.508 h.

# 16815

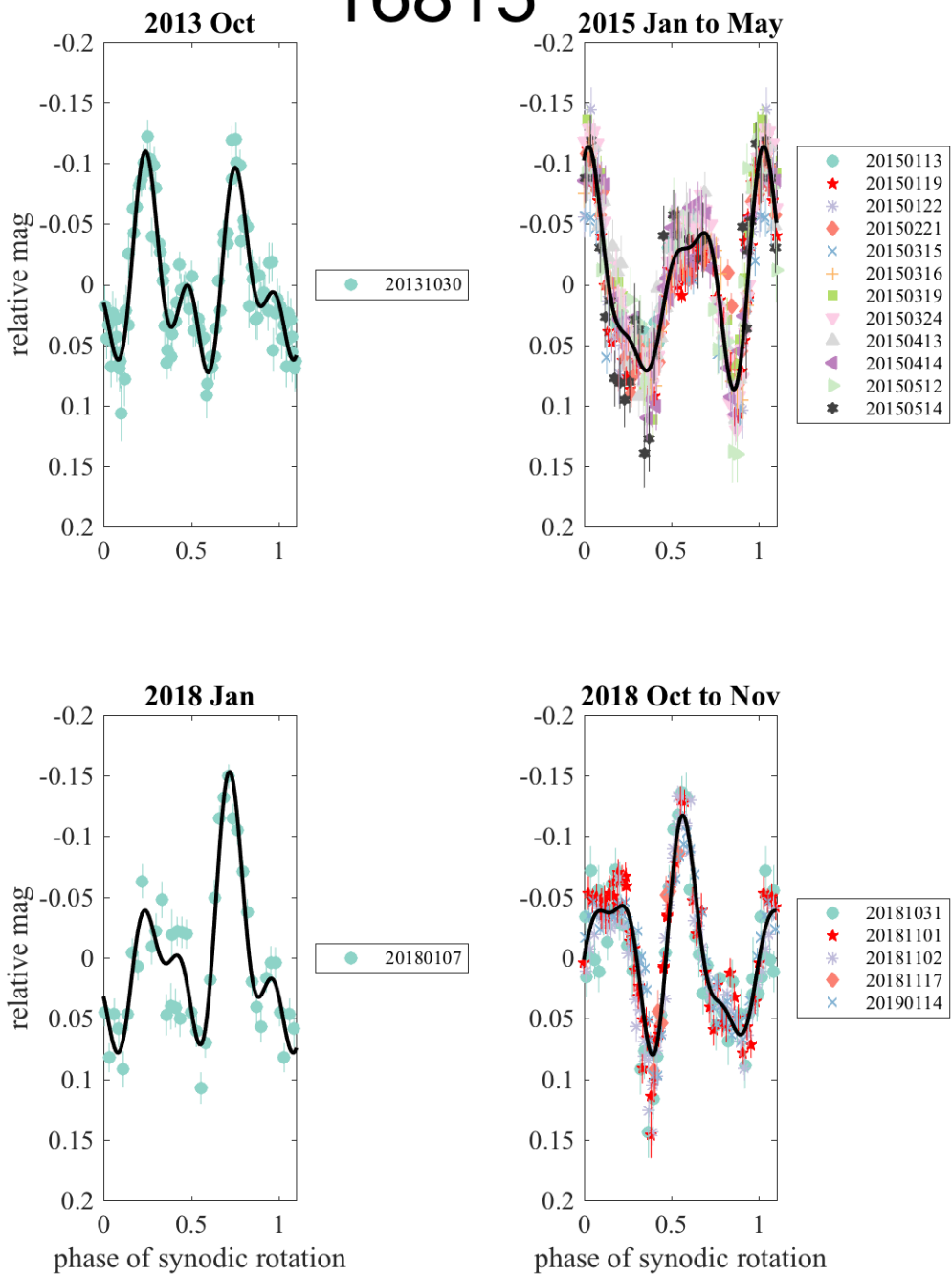

Fig. A8: Folded lightcurves of 16815 1997 EA9 using period of 2.9176 h.

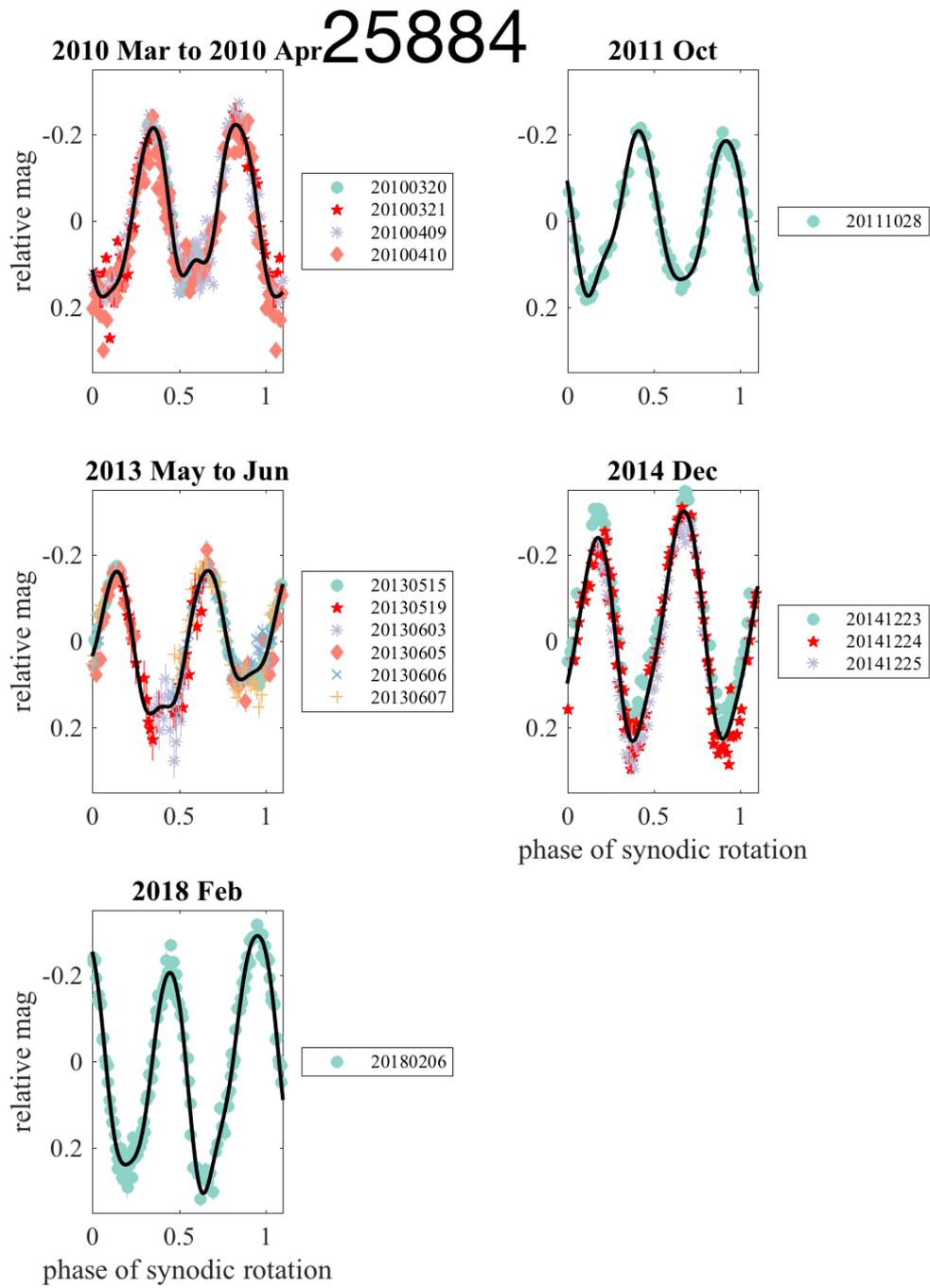

Fig. A9: Folded lightcurves of 25884 Asai using period of 4.9169 h.

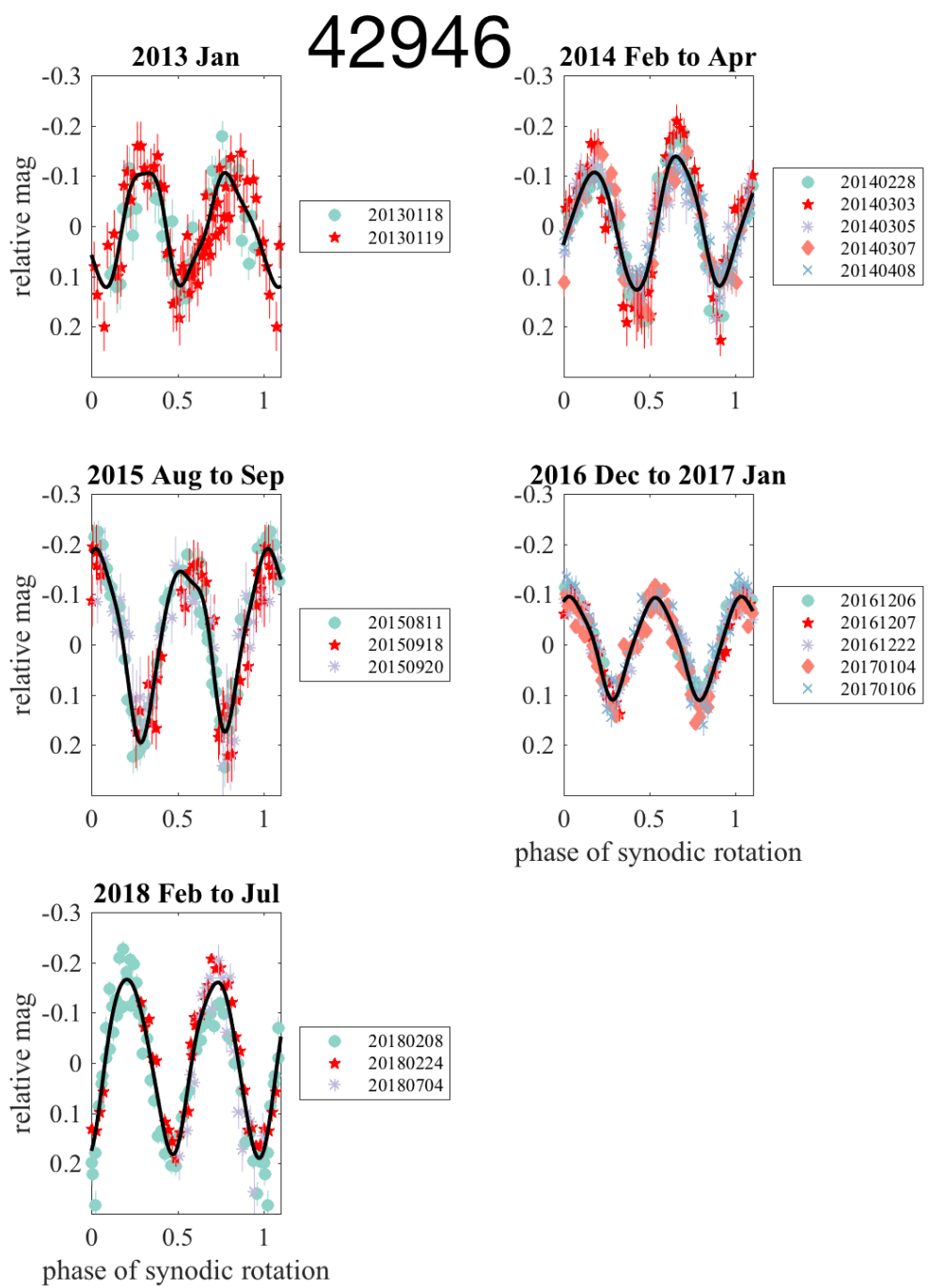

Fig. A10: Folded lightcurves of 42946 1999 TU95 using period of 3.4081 h.

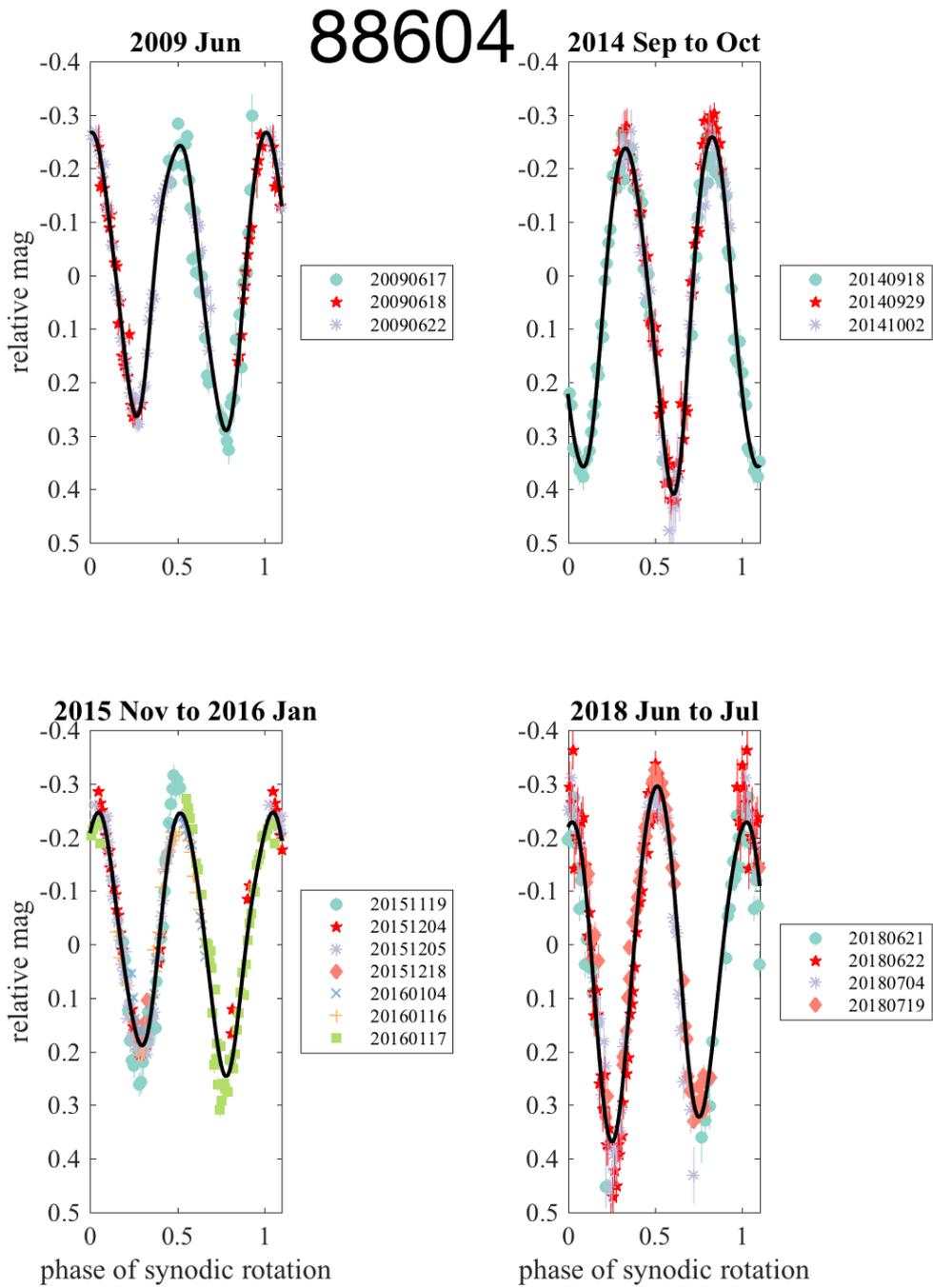

Fig. A11: Folded lightcurves of 88604 2001 QH293 using period of 7.172 h.